\documentclass[floatfix,aps,prx,reprint,longbibliography]{revtex4-2}
\usepackage{blindtext}
\usepackage{lipsum}
\usepackage{amsmath}
\usepackage{graphicx}

\begin{document}
\title{Disentangling Losses in Tantalum Superconducting Circuits}
\author{Kevin D. Crowley$^{1,\dag}$, Russell A. McLellan$^{2,\dag}$, Aveek Dutta$^{2,\dag}$, Nana Shumiya$^2$, Alexander P. M. Place$^2$, Xuan Hoang Le$^2$, Youqi Gang$^2$, Trisha Madhavan$^2$, Nishaad Khedkar$^2$, Yiming Cady Feng$^2$, Esha A. Umbarkar$^1$, Xin Gui$^3$, Lila V. H. Rodgers$^2$, Yichen Jia$^4$, Mayer M. Feldman$^1$, Stephen A. Lyon$^2$, Mingzhao Liu$^4$, Robert J. Cava$^3$, Andrew A. Houck$^2$, Nathalie P. de Leon$^{2,}$}
\email{npdeleon@princeton.edu \\
$^\dag$These authors contributed equally.}
\affiliation{$^1$Department of Physics, Princeton University, Princeton, New Jersey, 08540}
\affiliation{$^2$Department of Electrical and Computer Engineering, Princeton University, Princeton, New Jersey, 08540}
\affiliation{$^3$Department of Chemistry, Princeton University, Princeton, New Jersey, 08540}
\affiliation{$^4$Center for Functional Nanomaterials, Brookhaven National Laboratory, Upton, New York, 11973}

\begin{abstract}
Superconducting qubits are a leading system for realizing large scale quantum processors, but overall gate fidelities suffer from coherence times limited by microwave dielectric loss. Recently discovered tantalum-based qubits exhibit record lifetimes exceeding 0.3~ms. Here we perform systematic, detailed measurements of superconducting tantalum resonators in order to disentangle sources of loss that limit state-of-the-art tantalum devices. By studying the dependence of loss on temperature, microwave photon number, and device geometry, we quantify materials-related losses and observe that the losses are dominated by several types of saturable two level systems (TLSs), with evidence that both surface and bulk related TLSs contribute to loss. Moreover, we show that surface TLSs can be altered with chemical processing. With four different surface conditions, we quantitatively extract the linear absorption associated with different surface TLS sources. Finally, we quantify the impact of the chemical processing at single photon powers, the relevant conditions for qubit device performance. In this regime we measure resonators with internal quality factors ranging from 5 to 15~$\times\ 10^6$, comparable to the best qubits reported. In these devices the surface and bulk TLS contributions to loss are comparable, showing that systematic improvements in materials on both fronts will be necessary to improve qubit coherence further.
\end{abstract}

\maketitle

\section{Introduction}
Superconducting qubits have been deployed in some of the most sophisticated quantum processors, enabling demonstrations of quantum error correction \cite{krinner_realizing_2022,gong_experimental_2022, sivak_real-time_2022, zhao_realization_2022, acharya_suppressing_2022}, quantum many body physics and entanglement dynamics \cite{mi_noise-resilient_2022, mi_time-crystalline_2022, andersen_observation_2022, satzinger_realizing_2021}, and quantum simulation \cite{kandala_hardware-efficient_2017}. Improvements in superconducting qubit coherence would help to enable large-scale quantum processors, potentially capable of executing useful tasks. Current superconducting qubits are limited by dielectric loss that is orders of magnitude higher than expected from bulk properties of the constituent materials \cite{read_precision_2022, de_leon_materials_2021, krupka_complex_1999, krupka_use_1999}. This high dielectric loss indicates that qubit relaxation likely originates from uncontrolled surfaces, interfaces, and contaminants. Tantalum qubits have recently been demonstrated to exhibit record lifetimes and coherence times exceeding 0.3~ms \cite{placeRodgers2021}, which has been reproduced with different fabrication methods \cite{wang_towards_2022} and substrates \cite{lozano_manufacturing_2022}, indicating that major advances can be enabled by materials discovery. Tantalum qubits have also recently been deployed to achieve break-even quantum error correction \cite{sivak_real-time_2022}, and further improvements in coherence could allow current processors and architectures to push beyond the threshold for fault tolerance \cite{acharya_suppressing_2022, teoh_dual-rail_2022}. The advantage of tantalum likely arises from its stoichiometric, kinetically-limited oxide and its chemical robustness, allowing for extensive device cleaning \cite{placeRodgers2021, mclellan_chemical_2023}. However, little is known about the remaining sources of loss that limit state-of-the-art tantalum devices.

Prior work in other material systems has focused on the role of parasitic two-level systems (TLSs) in decoherence and dissipation \cite{gao_jiansong_physics_2008}. TLSs were originally explored in the context of thermal transport in glasses \cite{phillips_two-level_1987, muller_towards_2019}, and are ubiquitous sources of loss and decoherence in myriad systems, including superconducting devices \cite{martinis_decoherence_2005,paladino_mathbsf1mathbsfitf_2014, de_graaf_direct_2017, klimov_fluctuations_2018, lisenfeld_electric_2019, mcrae_materials_2020, woods_determining_2019, christensen_anomalous_2019}, microwave kinetic inductance detectors \cite{gao_noise_2007, kumar_temperature_2008, gao_jiansong_physics_2008}, optomechanical cavities \cite{riviere_optomechanical_2011}, and acoustic resonators \cite{wollack_loss_2021, maccabe_nano-acoustic_2020}. However, the magnitude of the TLS contribution to device loss is difficult to quantitatively disentangle from other sources of loss, such as radiative losses \cite{sage_study_2011}, packaging \cite{huang_microwave_2021}, nonequilibrium quasiparticles \cite{serniak_hot_2018}, and non-saturable absorption \cite{braginsky_experimental_1987, gurevich_intrinsic_1991}.  This identification is complicated by the likelihood that there are multiple TLS sources in a given device, whose relative contribution may depend on device geometry, fabrication and cleaning procedures, and subtle material choices.

Here we quantitatively separate different contributions to microwave loss arising from TLSs, quasiparticles, and other channels by varying temperature and microwave photon number. We observe internal quality factor ($Q_{\textrm{int}}$) up to 2~$\times\ 10^8$ at high power, giving a large dynamic range that allows us to measure subtle sources of loss. We observe a non-monotonic temperature dependence in $Q_{\textrm{int}}$, with the low temperature behavior well-described by TLS loss. Furthermore, we can quantify surface and bulk TLS contributions by varying device geometry. We find that smaller devices are dominated by surface TLSs, while contributions from TLSs residing in the bulk become evident in larger devices. By treating the devices with a post-fabrication buffered oxide etch (BOE), we can decrease the surface TLS bath, and by comparing different surface treatments, we can quantitatively estimate the contribution of different material interfaces. Finally, we characterize the different components of TLS loss at single microwave photon powers as a proxy for qubit performance.

\section{Resonator fabrication and measurements}
We deposit 200~nm of Ta epitaxially on 300 or 500~$\mu$m thick sapphire substrates using DC magnetron sputtering at elevated temperatures to stabilize the BCC $\alpha$-phase \cite{face_nucleation_1987, SI}. In contrast to our prior work \cite{placeRodgers2021}, all films are $\langle 111 \rangle$ oriented, single crystal, with some films having a minority component of the $\langle 110 \rangle$ orientation. We pattern resonators using photolithography followed by metal etching, either using a selective wet chemical etch or dry etching in an inductively-coupled plasma reactive ion etching system. We then strip the photoresist and clean the devices using a piranha solution composed of 1:2 hydrogen peroxide in sulfuric acid (``native" surface). Finally, in order to chemically alter the tantalum surface, some devices are treated with either a 10:1 buffered oxide etch for 20 minutes (``BOE" surface), a 10:1 buffered oxide etch for 120 minutes (``long BOE" surface), or a refluxing mixture of 1:1:1 concentrated sulfuric, nitric, and perchloric acids (``triacid" surface).

The fabricated devices consist of either coplanar waveguide (CPW) quarter-wave resonators or lumped element (LE) resonators. We vary their sizes to achieve different surface participation ratios (SPR) (Figs. 1a,b), the fraction of the electric field energy residing in surface layers of the device \cite{wang_surface_2015,mcrae_materials_2020}. The CPW resonators are shorted transmission lines with characteristic impedances of 50 $\Omega$, and the LE resonators are LC oscillators with characteristic impedance of 300-400 $\Omega$. Multiple resonators are coupled to a single feedline and are designed to have different resonant frequencies between 4 and 8 GHz to allow for spectrally selective interrogation.

We characterize the losses in each resonator by measuring transmission through the feedline in a dilution refrigerator with a base temperature around 17~mK and scanning the frequency of the probe tone around the resonant frequency \cite{SI}. At the resonant frequency, the lineshape of the transmission dip reflects both the internal losses and the coupling to the feedline. We fit the lineshape to extract the internal quality factor, $Q_{\rm{int}}$ (Fig. 1c)  \cite{geerlings_improving_2013, SI}.

A common observation in superconducting circuits is that losses decrease with increasing microwave power, indicating that the losses are from saturable TLSs \cite{de_leon_materials_2021}. We observe similar power dependent loss in our devices (Fig. 1d), with low power $Q_{\rm{int}}$ ranging from 1~$\times\ 10^5$ to 1~$\times\ 10^7$ and high power $Q_{\rm{int}}$ ranging from 1~$\times\ 10^7$ to 2~$\times\ 10^8$ across different devices.

\begin{figure}
    \centering
    \includegraphics{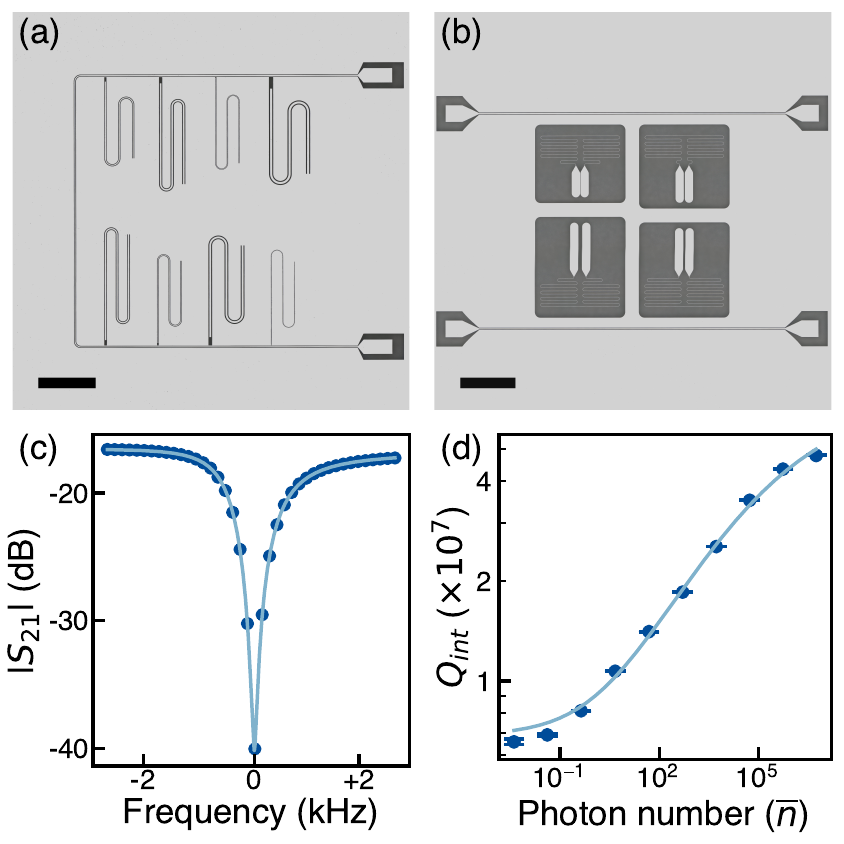}
    \caption{Measuring superconducting resonators. (a) Optical microscope image of a CPW resonator chip, consisting of 8 resonators with varying pitch capacitively coupled to a single RF feedline. Scale bar represents 1~mm. (b) Optical microscope image of an LE resonator chip, consisting of 4 resonators with varying capacitor spacing inductively coupled to a single RF feedline. Scale bar represents 1~mm. (c) Transmission spectrum of a single resonator, centered about the resonance frequency $f_0 = 4.484501$ GHz. Solid line indicates a fit to the data from which $Q_{\textrm{int}}$, $f_0$, and the coupling losses are extracted. (d) $Q_{\textrm{int}}$ as a function of the applied microwave power, expressed in terms of the average intracavity photon number. Solid line indicates a fit to the data based on saturation of TLSs and an additional power-independent loss.}
    \label{fig:fig1}
\end{figure}

\begin{figure}
    \centering
    \includegraphics{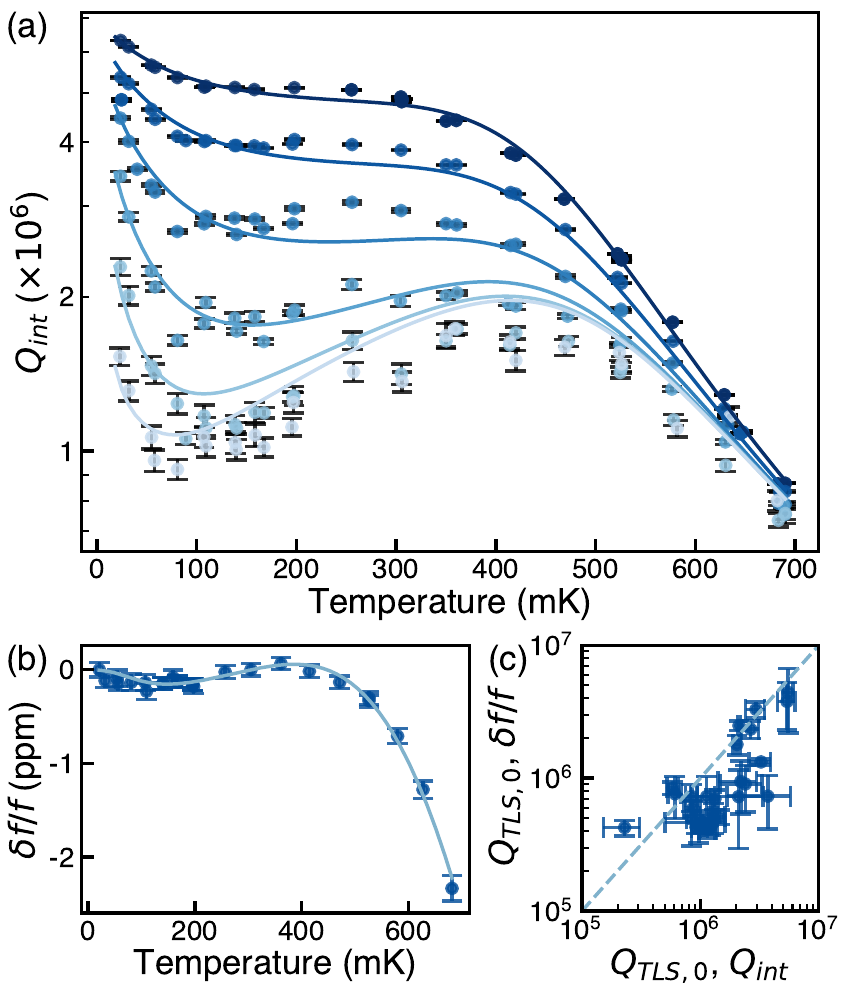}
    \caption{Parametrizing losses. (a) Internal quality factor $Q_{\textrm{int}}$ as a function of applied microwave power and temperature for a characteristic resonator. Color shade indicates the applied power measured at the output of the network analyzer, spaced 10~dB apart from the highest power of -10~dBm (darkest blue) to the lowest power of -60~dBm (lightest blue). The traces are well separated at low temperatures, and then collapse together and fall exponentially at high temperatures. The characteristic shape of the curves is fit to a model incorporating TLS loss and equilibrium quasiparticles. Solid lines show the best fit to the dataset. (b) Shift in the resonant frequency with temperature relative to the base temperature center frequency for a representative device. Solid line represents a fit to the data. (c) Comparison between estimates of $Q_{\textrm{TLS,0}}$ extracted from two independent measurements, the power and temperature dependence of $Q_{\textrm{int}}$ and the temperature dependence of the frequency. The dashed line is a guide to the eye showing the case where $Q_{\textrm{TLS,0}}$ is equal for both measurements. Only 26 devices are shown in this plot because most measurements were optimized for measuring $Q_{\textrm{int}}$, and did not have enough data to extract $Q_{\textrm{TLS,0}}$ from the fractional frequency shift with high confidence.}
    \label{fig:fi32}
\end{figure}

Different sources of loss can be distinguished by their power and temperature dependence. In order to further disentangle different physical mechanisms for loss, we characterize resonator losses over a wide range of temperatures and microwave powers (Fig. 2a). The full power and temperature dependence is well-described by a model that incorporates three sources of loss: TLSs ($Q_{\rm{TLS}}$), equilibrium quasiparticles ($Q_{\rm{QP}}$), and a separate power- and temperature-independent loss channel that limits $Q_{\rm{int}}$ at the highest microwave powers ($Q_{\rm{other}}$). We fit the full dataset using the following model:
\begin{equation}
    \frac{1}{Q_{\textrm{int}}} = \frac{1}{Q_\textrm{TLS}(\overline{n}, T)} + \frac{1}{Q_\textrm{QP}(T)} + \frac{1}{Q_\textrm{other}}.
\end{equation}
The TLS and quasiparticle losses are parametrized by \cite{gao_jiansong_physics_2008}:
\begin{equation}
    Q_\textrm{TLS}(\overline{n}, T) = Q_{\textrm{TLS,0}} \frac{\sqrt{1 + \left( \frac{\overline{n}^{\beta_2}}{D T^{\beta_1}} \right) \tanh \left( \frac{\hbar \omega}{2 k_B T} \right) }}{\tanh \left( \frac{\hbar \omega}{2 k_B T} \right) }
\end{equation}
and
\begin{equation}
    Q_\textrm{QP}(T) = Q_{\textrm{QP,0}} \frac{e^{\Delta_0 /k_B T}}{\sinh \left( \frac{\hbar \omega}{2 k_B T} \right) K_0 \left( \frac{\hbar \omega}{2 k_B T} \right)},
\end{equation}
where $\omega$ is the center angular frequency of the resonator; $T$ is the temperature; $\overline{n}$ is the intracavity photon number; $Q_\textrm{TLS,0}$ is the inverse linear absorption from TLSs; $D$, $\beta_1$ \cite{burnett_evidence_2014}, and $\beta_2$ \cite{wang_improving_2009} are parameters characterizing TLS saturation; $Q_{\textrm{QP,0}}$ is the inverse linear absorption from quasiparticles; $\Delta_0$ is the superconducting gap ($\Delta_0 = 1.764 k_B T_c$); $T_c$ is the superconducting critical temperature of the film; $K_0$ is the zeroth order modified Bessel function of the second kind;  $k_B$ is the Boltzmann constant; and $\hbar$ is the reduced Planck constant. There are seven free fit parameters: $Q_\textrm{TLS,0}$, $D$, $\beta_1$, $\beta_2$,  $Q_{\textrm{QP,0}}$, $T_c$, and $Q_{\textrm{other}}$. 

This model gives rise to three separate regimes in the data. At high temperatures above 500~mK, $Q_{\textrm{int}}$ exhibits a weak power dependence and decreases exponentially with temperature. This behavior is consistent with equilibrium quasiparticle loss as the temperature becomes an appreciable fraction of the superconducting critical temperature for $\alpha$-Ta \cite{gao_jiansong_physics_2008}. At intermediate temperatures, 100-500~mK, $Q_{\rm{int}}$ increases with temperature by around a factor of two for the lowest microwave powers, with little variation with temperature at the highest microwave powers, consistent with thermal saturation of TLSs, where the characteristic temperature is given by the resonator frequency \cite{gao_jiansong_physics_2008}. At the lowest temperatures, there is an apparent 1/$T$ dependence of $Q_{\textrm{int}}$. This behavior is consistent with a decreasing TLS coherence time with increasing temperature and subsequent increase of TLS saturation power \cite{gao_jiansong_physics_2008, quintana_characterization_2014}. In our devices, the saturation power at base temperature for the TLS bath can be as low as 0.01 photons.

At base temperature and at the powers relevant for transmon operation where the average intracavity photon number is $\overline{n}=1$, the dominant source of loss is $1/Q_\textrm{{TLS}}$. We focus on $Q_\textrm{{TLS,0}}$ as a parameter that captures linear absorption due to TLSs and therefore reveals differences in the materials under different fabrication conditions. We can check that $Q_\textrm{{TLS,0}}$ is a robust parameter by independently measuring the temperature-dependent shift in the frequency of the resonator (Fig. 2b). The resonance frequency shifts because of the change in the real part of the dielectric constant arising from losses associated with the entire spectral distribution of the TLS bath \cite{pappas_two_2011}, as well as losses induced by quasiparticles. The frequency shift is given by \cite{gao_jiansong_physics_2008}:

\begin{equation} \label{eq:freqShift_total}
    \frac{\delta f(T)}{f_0} = \left( \frac{\delta f(T)}{f_0} \right)_\textrm{TLS} + \left( \frac{\delta f(T)}{f_0} \right)_\textrm{QP},
\end{equation}
where $f_0$ is the center frequency of the resonator at zero temperature and $\delta f$ is the difference in the center frequency of the resonator at nonzero temperature.

The TLS and quasiparticle contributions to the frequency shift are given by \cite{gao_jiansong_physics_2008}:
\begin{equation}
    \begin{split}
        \left( \frac{\delta f (T)}{f_0} \right)_\textrm{TLS} = & \frac{1}{\pi Q_\textrm{TLS,0}} \textrm{Re} \left[ \Psi \left( \frac{1}{2} + i \frac{\hbar \omega}{2 \pi k_B T} \right) \right.  \\
        & \left. - \ln \left( \frac{\hbar \omega}{2 \pi k_B T} \right) \right] 
    \end{split}
\end{equation}
and
\begin{equation}
    \begin{split} \label{eq:freqShift_QP}
        \left( \frac{\delta f (T)}{f_0} \right)_\textrm{QP} & = -\frac{\alpha}{2} \left( 1 - \vphantom{\sqrt{ \frac{\sigma_1 (T, \omega)^2}{\sigma_1 (0, \omega)^2} }} \right. \\
        & \left. \sin \left( \phi (T, \omega) \right) \sqrt{ \frac{\sigma_1 (T, \omega)^2 + \sigma_2 (T, \omega)^2}{\sigma_1 (0, \omega)^2 + \sigma_2 (0, \omega)^2} } \right),
    \end{split}
\end{equation}
where $\Psi$ is the complex digamma function; $\sigma_1$ and $\sigma_2$ are the real and imaginary parts of the complex conductivity; $\phi$ is the phase between the real and imaginary parts of the complex conductivity; and $\alpha$ is the kinetic inductance fraction \cite{SI}. The three free fit parameters are $Q_\textrm{TLS,0}$, $T_c$, and $\alpha$.

We compare the extracted $Q_\textrm{TLS,0}$ from the two measurements (Fig. 2c) and find that they agree on average to within a factor of 2.6$\sigma$. We note that our measurements were optimized for measuring $Q_{\textrm{int}}$ rather than $\delta{}f/f_0$. As $\delta{}f/f_0$ is not sensitive to the applied microwave power, we have a factor of 5 to 10 fewer data points with which to fit $\delta{}f/f_0$ than $Q_{\textrm{int}}$. We conclude that while the difference between the two measurements is statistically significant, $Q_\textrm{TLS,0}$ is a robust parameter that forms a quantitative basis of comparison across devices when fitted from $Q_{\textrm{int}}$ data.

\begin{figure*}[htbp]
    \centering
    \includegraphics{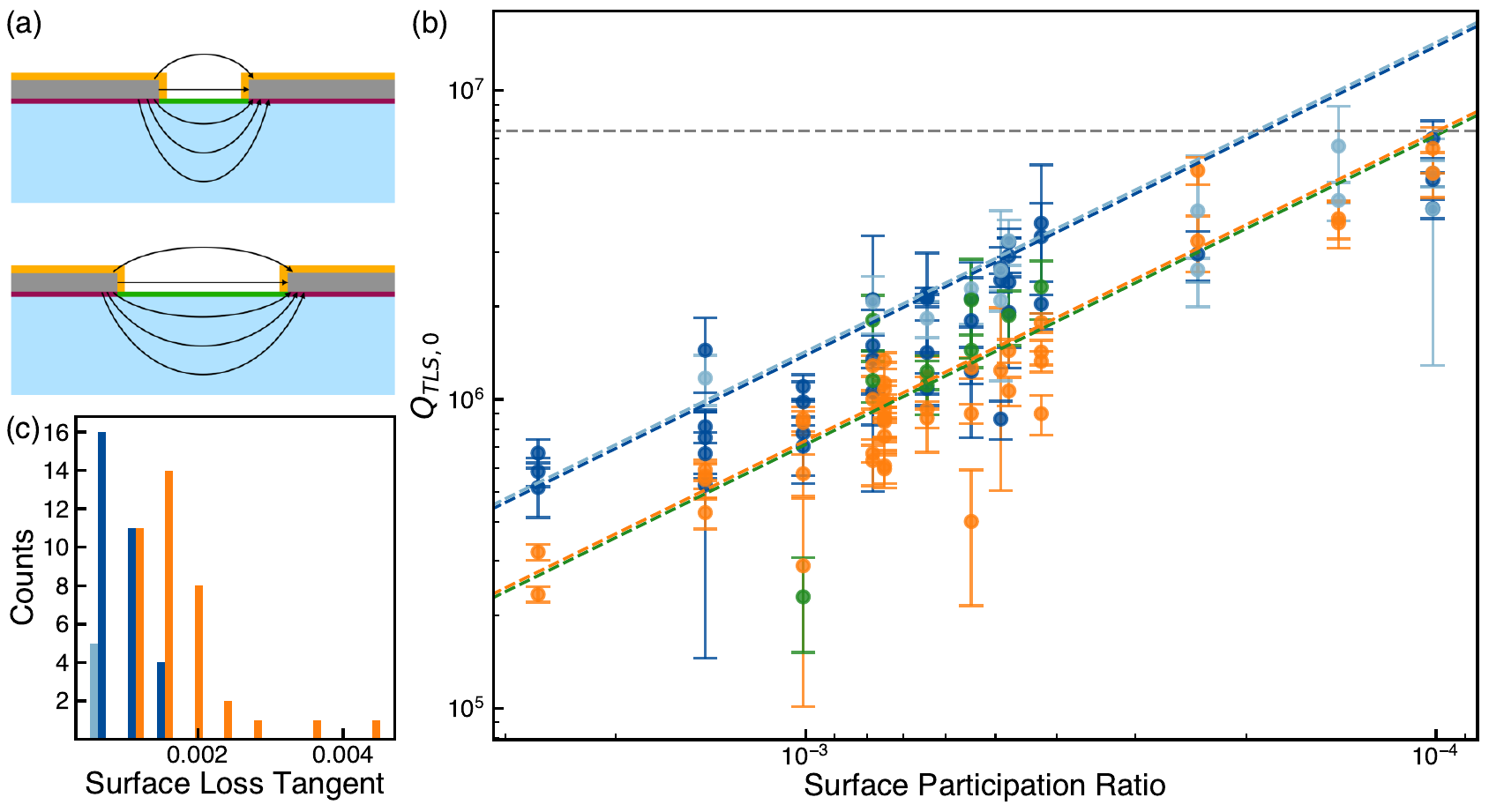}
    \caption{Dependence of loss on SPR. (a) Cartoon cross-section illustrating the dependence of SPR on device geometry. As the distance between capacitor pads (gray) increases, the fraction of the electric field (black arrows) energy overlapping with a thin layer at the three interfaces, metal-air (yellow), metal-substrate (purple), and substrate-air (green), decreases. The fraction of the electric field energy in the sapphire substrate (blue) does not strongly depend on the distance between capacitor pads. (b) Dependence of the extracted $Q_{\textrm{TLS,0}}$ from $Q_{\textrm{int}}$ measurements on SPR. For the highest SPR devices $Q_{\textrm{TLS,0}}$ exhibits linear scaling with SPR, but for lower SPR the $Q_{\textrm{TLS,0}}$ saturates, indicating that there are both surface and bulk TLS baths. Comparing the data for four different surface conditions, native (orange), BOE (dark blue), long BOE (light blue), and triacid (green) allows for the estimate of surface loss tangents for each surface: $\tan \delta_{\textrm{surface,native}}$ (dashed orange line), $\tan \delta_{\textrm{surface,BOE}}$ (dashed dark blue line), $\tan \delta_{\textrm{surface,long BOE}}$ (dashed light blue line), and $\tan \delta_{\textrm{surface,triacid}}$ (dashed green line), as well as the bulk substrate loss tangent $\tan \delta_{\textrm{bulk}}$ (dashed grey line). (c) Histogram of surface loss tangents for native (orange), BOE (dark blue) and long BOE (light blue) CPW devices, showing that BOE and long BOE treatments result in surface loss tangents that are a factor of 1.89 times and 1.96 times lower than the native surface on average.}
    \label{fig:fig2}
\end{figure*}

\section{Parametrizing sources of loss}

TLSs that cause loss can occur in many different materials in the same device: surface oxides, surface contamination, the exposed sapphire surface, the tantalum-sapphire interface, the bulk of the sapphire, and other elements related to packaging. In order to identify the location and origin of TLS loss, we fabricated 24 chips containing a total of 105 devices with varying geometry and surface conditions, and performed temperature and power dependent loss measurements to extract $Q_{\textrm{TLS,0}}$. Varying the device geometry changes the SPR (Fig. 3a). By modeling interfaces as dielectrics with an assumed standard thickness (3~nm) and permittivity ($\epsilon$=10), we can compute the fraction of electric field energy that overlaps with the interfaces of a device for a given electromagnetic mode, the SPR  \cite{wang_surface_2015, SI}. 

For the CPW resonators, we tune the SPR by tuning the pitch of the shorted CPW transmission line. Fixing the impedance to be 50~$\Omega$ constrains the ratio between the centerpin width and the gap width \cite{simons_coplanar_2008}, so that the centerpin also increases in width as the SPR is reduced. For the LE resonators, we tune SPR by changing the spacing and size of the capacitor pads, where larger spacings and pads correspond to lower SPR. Across both types of devices, we vary the SPR by a factor of 30 \cite{SI}.

The extracted $Q_\textrm{{TLS,0}}$ increases with decreasing SPR (Fig. 3b). The trend is approximately linear and then plateaus for low SPR, below 3~$\times\ 10^{-4}$. We model this SPR dependence as arising from two different components, a surface-related TLS loss that scales with SPR, and a bulk TLS loss that is SPR-independent. The losses can be parametrized as a loss tangent, which is the ratio of the imaginary and real components of the dielectric constant, $\tan \delta = \textrm{Im}(\epsilon) / \textrm{Re}(\epsilon)$. The apparent surface loss tangent varies across the four surface treatments. Fitting these two components to the full dataset across all 105 devices yields surface loss tangents of $\tan \delta_{\textrm{surface,BOE}} = (7.2 \pm 0.6) \times 10^{-4}$, $\tan \delta_{\textrm{surface,long BOE}} = (7 \pm 1) \times 10^{-4}$, $\tan \delta_{\textrm{surface,native}} = (13.6 \pm 0.6) \times 10^{-4}$, and $\tan \delta_{\textrm{surface,triacid}} = (14 \pm 3) \times 10^{-4}$ for the BOE, long BOE, native, and triacid surface treatments, respectively. The fitted bulk loss tangent is $\tan \delta_{\textrm{bulk}} = (1.5 \pm 0.2) \times 10^{-7}$, which is an order of magnitude higher than recent bulk measurements on the same substrates \cite{read_precision_2022}. This indicates that the bulk loss we observe is dominated by a surface damage layer within 50~$\mu$m of the surface, rather than a uniform loss tangent throughout the bulk \cite{SI}.

We study the correlation between loss tangent and tantalum oxide thickness after the four different surface treatments to localize the source of surface-related TLS loss. The native oxide is an approximately 3~nm thick, kinetically-limited, stoichiometric oxide that is remarkably robust to chemical processing, as measured using X-ray photoelectron spectroscopy and transmission electron microscopy \cite{placeRodgers2021, mclellan_chemical_2023, SI}. In \cite{mclellan_chemical_2023}, we observe a reduction in oxide thickness after BOE treatment to approximately 2.4~nm, while after triacid treatment, the oxide grows to nearly 6~nm. We correlate results from \cite{mclellan_chemical_2023} with lab-based XPS results to estimate that the total oxide thickness after the long BOE treatment is 1.5~nm~$\pm$~0.3~nm \cite{SI}. The long BOE and BOE treated devices exhibit 1.96 and 1.89 times higher $Q_\textrm{{TLS,0}}$ than the devices with a native surface (Figs. 3b,c). Since the tantalum oxide layer is amorphous, and it is thinner after either BOE treatment, a likely hypothesis for the origin of TLS loss is the oxide layer. However, we observe that the triacid-treated samples have similar values of $Q_\textrm{{TLS,0}}$ to those of the native samples, despite their thicker oxide. Therefore, the TLS loss is not proportional to the volume of the oxide, possibly because another bath of TLSs decreases to compensate the additional oxide-related loss in the triacid treated samples.

One possible difference among the samples is that the triacid and BOE treatments are highly effective at removing residual hydrocarbon contamination from fabrication, resulting in a reduction in the parasitic hydrocarbon TLS loss commensurate with the increased oxide loss from triacid treatment. Treatment in BOE removes some surface oxide, therefore any contamination on that surface should be removed concurrently; similarly, the triacid treatment has previously been shown to be strongly oxidizing and effective at removing hydrocarbons \cite{sangtawesin_origins_2019}. We therefore model the observed surface loss tangents for the three surface conditions as arising from three components: a hydrocarbon component at the metal-air interface, an oxide-related loss tangent that is proportional to the oxide thickness, and a component related to the metal-substrate and substrate-air interfaces \cite{SI}:

\begin{equation} \label{eq:lossModel}
    \begin{split}
      \frac{1}{Q_{\textrm{TLS,0}}} = &\ p_{\textrm{MS}} \tan \delta_{\textrm{surface},i} \\ = & \  p_{\textrm{MA}}  \left( \left(
\frac{t_{\textrm{TaOx},i}}{t_0} \right) \tan \delta_{\textrm{TaOx}} + \gamma_i \tan \delta_{\textrm{HC}} \right) \\ & + p_{\textrm{MS}} \tan \delta_\textrm{MS} + p_{\textrm{SA}} \tan \delta_\textrm{SA},
   \end{split}
\end{equation}
where $\tan \delta_{\textrm{surface},i}$ is the loss tangent for condition $i$, $t_0$ is the assumed thickness of the substrate-air and metal-substrate interfaces used to simulate participation ratios (3~nm), $t_{\textrm{TaOx},i}$ is the oxide thickness of treatment $i$, $\gamma_i \in \{0,1\}$ is a factor determining if hydrocarbon loss is considered for surface condition $i$, and the subscripts MA, SA, MS, and HC refer to metal-air, substrate-air, metal-substrate, and hydrocarbons respectively. 

To estimate the different components of the loss, we assume that the hydrocarbon loss is completely eliminated after any of the triacid or BOE treatments ($\gamma_i$ = 0) and is present only for the native condition ($\gamma_\textrm{native} =1$). By quantitatively comparing the extracted loss tangents for our four conditions with Equation~\ref{eq:lossModel} \cite{SI}, we calculate a putative rescaled hydrocarbon-related loss tangent for an assumed standard 3~nm thick interface, $\frac{p_{\textrm{MA}}}{p_{\textrm{MS}}}\tan \delta_{\textrm{HC}} = (4.9 \pm 0.5) \times 10^{-4}$, and an intrinsic loss tangent for the tantalum oxide, $\tan \delta_{\textrm{TaOx}} = (5 \pm 1) \times 10^{-3}$. When rescaled to account for the difference in participation ratios of different surfaces,  $\frac{p_{\textrm{MA}}}{p_{\textrm{MS}}}\tan \delta_{\textrm{TaOx}}=(5 \pm 1) \times 10^{-4}$.

Assuming that the triacid and two BOE treatments do not affect the metal-substrate and substrate-air interfaces, our model also provides an estimate for the loss contributions of those two interfaces, and finds that they give a combined value of $\tan \delta_{\textrm{MS}}+\frac{p_{\textrm{SA}}}{p_{\textrm{MS}}}\tan \delta_{\textrm{SA}} = (4 \pm 1) \times 10^{-4}$ \cite{SI}. The rescaled loss tangents are all comparable, indicating that all surfaces play a critical role in determining overall loss. Other possible models for the effects of surface treatment that we have ruled out based on the data are detailed in the Supplemental Material \cite{SI}.

Our data rules out a model for the TLS loss that is purely extensive in the oxide thickness. Here we have hypothesized that a second bath residing in fabrication-related contaminant hydrocarbons can account for the difference. However, there are other possible microscopic models, such as the possibility that a single chemical species or suboxide is responsible for all the TLS loss in the oxide, and the native and triacid samples have equal amounts of that species despite the large difference in total oxide thickness. Testing such hypotheses would require the measurement of many more surface conditions that independently vary each candidate TLS component.

While $Q_{\textrm{TLS,0}}$ parametrizes the linear absorption in the device, for transmon operation the steady state photon occupation will be around $\overline{n}=1$, and we observe that the saturation power for the TLS bath can be a small fraction of a single photon. For the largest devices at base temperature, $Q_{\textrm{TLS}}(\overline{n}=1)$ ranges from 5 to 15~$\times\ 10^6$ (Fig. 4), in line with state-of-the-art qubits \cite{placeRodgers2021,wang_towards_2022}. 

Examining the dependence of $Q_{\textrm{TLS}}(\overline{n}=1)$ on SPR reveals that BOE treatment also leads to an improvement. The SPR dependence is roughly linear despite the nonlinear dependence of $Q_{\textrm{TLS}}$ on microwave photon number. By performing the same analysis on the SPR scaling of $Q_{\textrm{TLS}}(\overline{n}=1)$ we extract apparent loss tangents at single photon powers for the four surface conditions, $\tan \delta_{\textrm{surface,BOE}}(\overline{n}=1) = (6.6 \pm 0.4) \times 10^{-4}$, $\tan \delta_{\textrm{surface,long BOE}}(\overline{n}=1) = (7 \pm 2) \times 10^{-4}$, $\tan \delta_{\textrm{surface,native}}(\overline{n}=1) = (11.7 \pm 0.5) \times 10^{-4}$, and $\tan \delta_{\textrm{surface,triacid}}(\overline{n}=1) = (11 \pm 1) \times 10^{-4}$. Again, for the largest devices surface and bulk loss are comparable, indicating that improvements in both will be required to improve overall device performance.

\begin{figure}[htbp]
    \centering
    \includegraphics{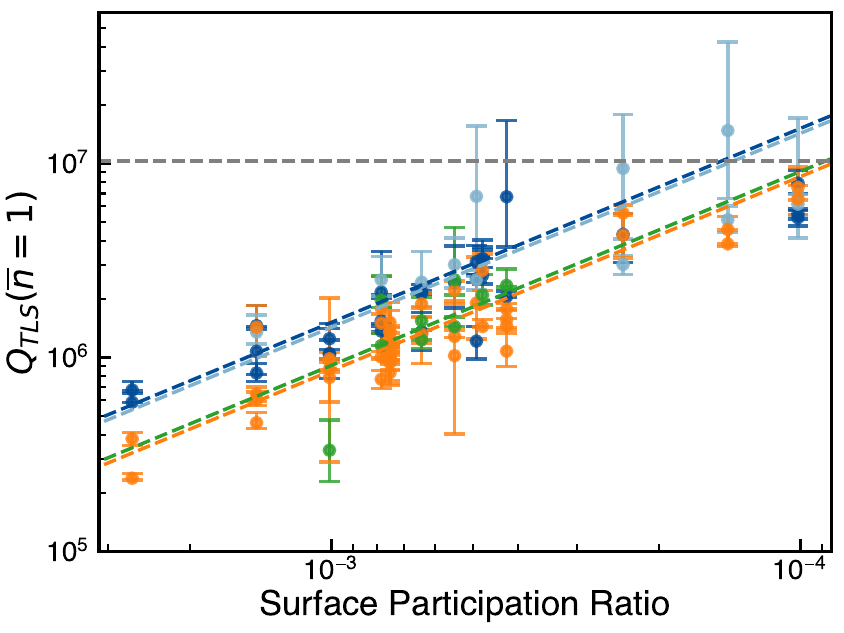}
    \caption{Dependence of $Q_{\textrm{TLS}}(\overline{n}=1)$ on SPR. $Q_{\textrm{TLS}}(\overline{n}=1)$ is calculated at base temperature. Data from four different surface conditions, native (orange), BOE (dark blue), long BOE (light blue), and triacid (green) are shown. While $Q_{\textrm{TLS}}$ is a nonlinear function of $\overline{n}$, we nevertheless observe a roughly linear dependence between $Q_{\textrm{TLS}}(\overline{n}=1)$ and SPR. We fit an apparent loss tangent to each surface condition: $\tan \delta_{\textrm{surface,native}}(\overline{n}=1)$ (dashed orange line), $\tan \delta_{\textrm{surface, BOE}}(\overline{n}=1)$ (dashed dark blue line), $\tan \delta_{\textrm{surface, long BOE}}(\overline{n}=1)$ (dashed light blue line), and $\tan \delta_{\textrm{surface,triacid}}(\overline{n}=1)$ (dashed green line), as well as the bulk substrate loss tangent $\tan \delta_{\textrm{bulk}}(\overline{n}=1)$ (dashed grey line). Data are calculated from Equation \ref{eq:lossModel} with errors propagated from errors in fit parameters. The error bars are truncated at the lower end by $Q_{\textrm{TLS,0}}$.}
    \label{fig:fig4}
\end{figure}

\section{Conclusion}
We observe that state-of-the-art tantalum devices are limited by TLS loss. Using systematic measurements and parametrization of losses in superconducting circuits, we have shown that there are multiple sources of TLSs: a surface-related TLS bath associated with the tantalum oxide that can be reduced by around a factor of two with BOE treatment, and a substrate-related TLS bath. Furthermore, the surface-related TLS loss is not extensive in the oxide volume, indicating that there may be at least one additional TLS bath, such as fabrication-related hydrocarbon contamination. Critically, each of these components is of similar magnitude for state-of-the-art devices, and future improvements in superconducting qubits will require material improvements that address all of these sources of loss. Two natural avenues to pursue based on our findings would be to passivate the Ta surface to avoid oxide formation entirely, and to study subsurface damage from polishing and surface processing in sapphire substrates. 

The observed temperature dependence also points to two paths for improving the performance of superconducting qubits: reducing the density of TLSs, and improving the coherence time of the TLS bath. Our ongoing work includes studying the dynamics of the TLS bath using pump-probe spectroscopy \cite{sage_study_2011} and other time-domain methods \cite{heidler_non-markovian_2021}.

Correlations between the measurements presented here and direct materials spectroscopy may identify atomistic origins of TLS loss. For example, the losses in the tantalum oxide could arise from particular suboxides or interface states, and detailed chemical profiling using X-ray photoelectron spectroscopy could elucidate the particular chemical species responsible for TLS loss \cite{premkumar_microscopic_2021, mclellan_chemical_2023}. More broadly, the parameterization presented here isolates and identifies the material-related loss, thereby enabling quantitative comparisons among different material systems, such as new superconducting metals \cite{chang_improved_2013} and metal heterostructures, alternative substrates such as high purity silicon \cite{creedon_high_2011}, and different fabrication and post-processing techniques.

\begin{acknowledgements}
We thank Rob Schoelkopf, Suhas Ganjam, Michel Devoret, Steve Girvin, Will Oliver, Chen Wang, Mike Hatridge, and Jeff Thompson for helpful discussions. This work was primarily supported by the U.S. Department of Energy, Office of Science, National Quantum Information Science Research Centers, Co-design Center for Quantum Advantage (C2QA) under contract number DE-SC0012704. Film characterization and processing was partly supported by the National Science Foundation (RAISE DMR-1839199). The authors acknowledge the use of Princeton’s Imaging and Analysis Center (IAC), which is partially supported by the Princeton Center for Complex Materials (PCCM), a National Science Foundation (NSF) Materials Research Science and Engineering Center (MRSEC; DMR-2011750), as well as the Princeton Micro/Nano Fabrication Laboratory. We also acknowledge MIT Lincoln Labs for supplying a traveling wave parametric amplifier.
\end{acknowledgements}

\bibliography{resonatorPaperBibliography} 

\end{document}


\title{Supplemental Material for: ``Disentangling Losses in Tantalum Superconducting Circuits"}
\author{Kevin D. Crowley$^{1,\dag}$, Russell A. McLellan$^{2,\dag}$, Aveek Dutta$^{2,\dag}$, Nana Shumiya$^2$, Alexander P. M. Place$^2$, Xuan Hoang Le$^2$, Youqi Gang$^2$, Trisha Madhavan$^2$, Nishaad Khedkar$^2$, Yiming Cady Feng$^2$, Esha A. Umbarkar$^1$, Xin Gui$^3$, Lila V. H. Rodgers$^2$, Yichen Jia$^4$, Mayer M. Feldman$^1$, Stephen A. Lyon$^2$, Mingzhao Liu$^4$, Robert J. Cava$^3$, Andrew A. Houck$^2$, Nathalie P. de Leon$^{2,}$}
\email{npdeleon@princeton.edu, \\
$^\dagger$These authors contributed equally.}
\affiliation{$^1$Department of Physics, Princeton University, Princeton, New Jersey, 08540}
\affiliation{$^2$Department of Electrical and Computer Engineering, Princeton University, Princeton, New Jersey, 08540}
\affiliation{$^3$Department of Chemistry, Princeton University, Princeton, New Jersey, 08540}
\affiliation{$^4$Center for Functional Nanomaterials, Brookhaven National Laboratory, Upton, New York, 11973}

\date{\today}
\maketitle
\label{Sec:SI}

\setcounter{figure}{0}
\setcounter{section}{0}
\renewcommand{\thefigure}{S\arabic{figure}}
\renewcommand{\thetable}{S\Roman{table}}
\renewcommand{\thesection}{\Roman{section}}

\section{\label{SI_Setup} Supplementary experimental methods}

\subsection{Sample fabrication} \label{sec:fab}

3" diameter sapphire substrates are cleaned in a 2:1 H$_2$SO$_4$:H$_2$O$_2$ piranha solution for 20 mins, then rinsed in 3 cups of de-ionized water follwed by 1 cup of 2-propanol, and then blow dried in N$_2$. Then the sapphire substrate is loaded into a DC magnetron sputtering system (AJA Orion 8). The substrate is heated in-situ at 850\degree C before tantalum sputtering. The film deposition parameters were as follows: RF power of 250~W, Ar flow rate of 30 sccm, ambient pressure, temperature ramp rate 1\degree C/minute, and steady state temperature of 750\degree C, which results in a film growth rate of approximately 8~nm/minute. Post deposition, the tantalum films are confirmed to be predominantly 
 $\langle111\rangle$ orientation in the $\alpha$-phase using a Bruker D8 Advance X-ray Diffractometer. The deposited tantalum film is dehydration baked at 110\degree C and then cooled for about a minute on a metal plate. Following this, AZ1518 is spun on at 4000 rpm for 45 secs with a ramp rate rate of 1000 rpm/sec for an approximate resist thickness of 3 $\mu$m and soft baked at 95\degree C for 1 minute. The photoresist is patterned using a Heidelberg DL66+ laser writer with a 1.8 $\mu$m spot size with a 50\% attenuator, intensity setting of 30\% and focus offset setting of 10\%. The photoresist is developed in AZ300MIF solution for 90~s and rinsed in de-ionized water for 30~s. After development, the mask is hard baked at 110\degree C for 2 minutes and then cooled on a metal plate for 1 minute.

Using the patterned photoresist as a mask, we etched each device with one of three different etch types. One type is a wet chemical etch, 1:1:1 ratio of HF:HNO$_3$:H$_2$O (Transene Tantalum Etchant 111), in which a sample is swirled for 21 seconds before being rinsed in 3 cups of de-ionized water and 1 cup of 2-propanol, then blown dry in N$_2$. The second etch type is a chlorine-based dry chemical etch in an inductively-coupled plasma reactive ion etcher (PlasmaTherm Takachi). The etching parameters for the chlorine dry etch are as follows: ambient pressure of 5.4 mTorr, chlorine flow rate of 5 sccm, argon flow rate of 5 sccm, RF power of 500 W, and bias power of 50 W, which results in an etch rate of approximately 100 nm/min. The third etch type is a fluorine based dry etch, using the same reactive ion etcher as the chlorine etch, with parameters: ambient pressure 50~mTorr, CHF$_3$ flow rate 40 sccm, SF$_6$ flow rate 15 sccm, Ar$_3$ flow rate 10 sccm, RF power of 100~W, and bias power of 100~W.

After etching, the photoresist mask is stripped in a Remover PG bath at 80\degree C for 1 hour followed by rinsing in 2-propanol. The patterned Ta film is coated with hard-baked AZ1518 using the same parameters mentioned above to act as a protective layer for dicing. The wafer is diced (Advanced Dicing Technologies proVectus 7100 dicing saw) into 10 mm or 7 mm pieces, depending on the packaging used in the dilution refrigerator. Following dicing, the photoresist is stripped in a Remover PG bath at 80\degree C for 1 hour, followed by 2 minutes each sonication in toluene, acetone, and 2-propanol. Some chips were sonicated in methanol for 2 minutes between the acetone and 2-propanol sonication to remove zinc contamination. The chips are blown dry in N$_2$, and then cleaned in a 2:1 H$_2$SO$_4$:H$_2$O$_2$ piranha solution for 20 mins followed by rinsing in 3 cups of de-ionized water and 1 cup of 2-propanol and then blow dried in N$_2$. 

After fabrication, the samples are treated in BOE or triacid as detailed in Section IB. Then the chips are bonded to a PCB using an automatic wire bonder (Questar Q7800). We used two types of packages for our resonator chips. One comprises a Cu-plated PCB and a Cu puck and penny coated with 1 $\mu$m Aluminum. The second comprises a commercial microwave package (QDevil QCage.24) with an associated Au-plated PCB. The mounting of the package to the dilution refrigerator is described in Section \ref{sec:measurement}.

\subsection{Surface processing}

10:1 buffered oxide etch (BOE, Transene) is a mixture of 10 parts 40\% NH$_4$F solution to 1 part 49\% HF solution by volume. BOE treated samples were placed in buffered oxide etch at room temperature and were not agitated. After 20 minutes (``BOE" treatment) or 120 minutes (``long BOE" treatment), the samples were removed and triple rinsed in de-ionized water and 2-propanol before being blown dry in N$_2$.

The triacid treatment is 1:1:1 equal mix by volume of 95-98\% H$_2$SO$_4$, 70\% HNO$_3$, and 70\% HClO$_4$ solutions (all percentages by weight). We procured all solutions from SigmaAldrich (catalogue numbers: H$_2$SO$_4$ - 258105, HNO$_3$ - 225711, HClO$_4$ - 244252). The mixture is refluxed for 2 hours and then allowed to cool for 1 hour. After cooling, the sample was removed, triply rinsed in de-ionized water and 2-propanol before being blown dry in N$_2$.

\subsection{Measurement apparatus} \label{sec:measurement}

\begin{figure}[htbp]
    \centering
    \includegraphics[width=2.25in]{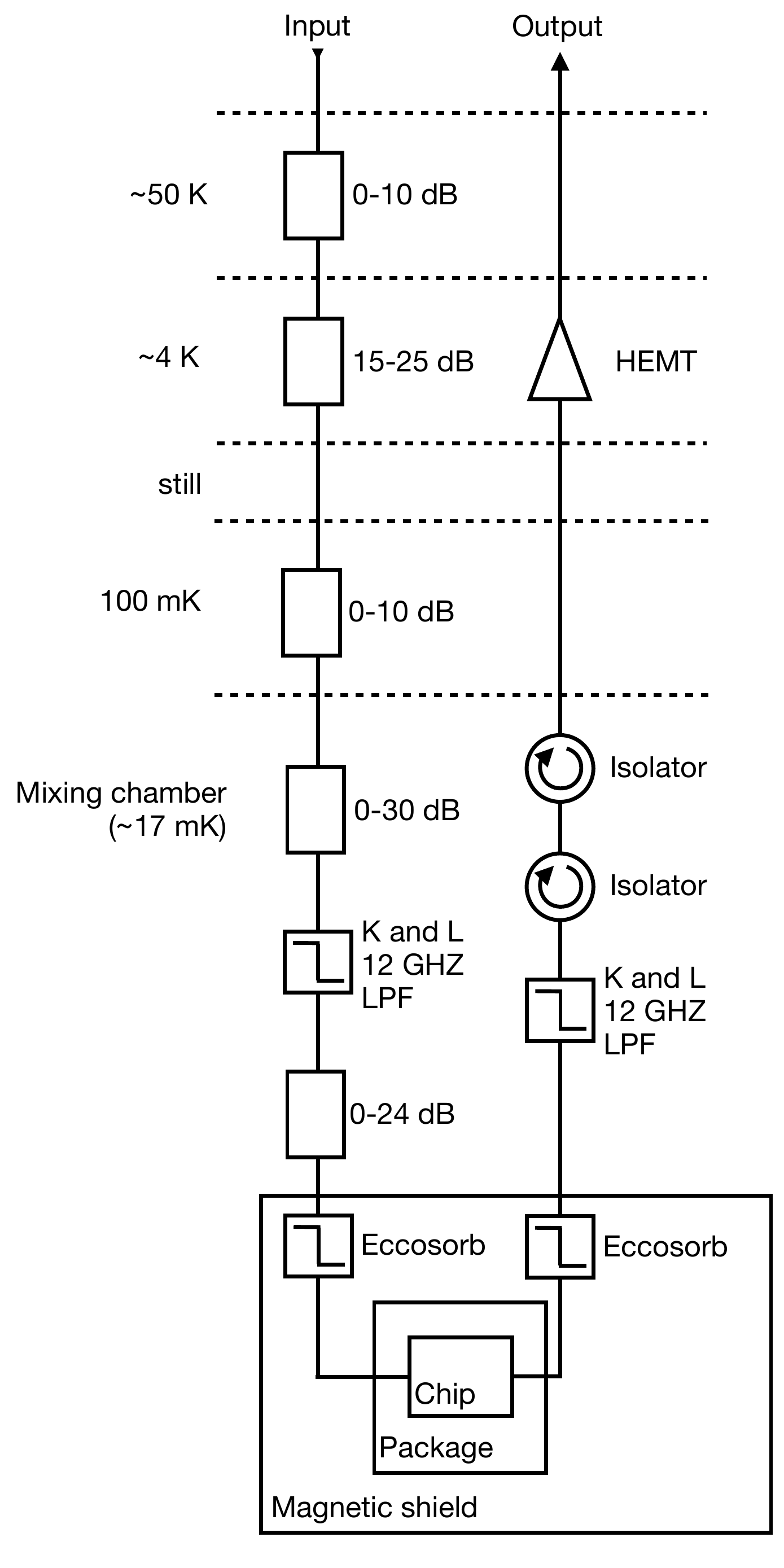}
    \caption{Wiring diagram for each of our measurement lines. Ranges of attenuation are given where the attenuation varied from line to line. The magnetic shields in our experiments varied between a QCan supplied by QDevil and a custom made mu-metal can. A traveling wave parametric amplifier (TWPA) was sometimes used on the output line in our experiments, and was placed in a separate magnetic shield and wired after the second isolator.}
    \label{fig:sup_fridge}
\end{figure}

All devices were measured in a BlueFors XLD dilution refrigerator with a base mixing chamber temperature of approximately 17~mK. There are four independent input lines and four corresponding output lines. A fridge diagram showing the layout for all four input and output lines is given in Figure \ref{fig:sup_fridge}.

Each input line has between 60 dB and 85 dB of attenuation from discrete cryogenic XMA attenuators (above mixing chamber, PN: 2082-604X-dB-CRYO) and cryogenic attenuators from Quantum Microwave (at mixing chamber, PNs: QMC-CRYOATTF-06 and  QMC-CRYOATTF-03), as well as attenuation from stainless steel coaxial transmission line cables, SMA connections, and insertion losses from filters. The total input line attenuation varies across the lines from 86.7 dB to 108.7 dB at resonator frequencies. Two types of low pass filter are used at the mixing chamber, a commercial filter from K\&L Microwave (PN: 6L250-00089) outside of the magnetic shield and an eccosorb filter placed inside of the magnetic shield. Two types of magnetic shield were used across our experiments. One type is a custom-fabricated can made of mu-metal, with which we used a custom made eccosorb filter with upper cutoff frequency approximately 8~GHz, and the other is a prototype product (QCan) from QDevil with which we used an eccosorb filter supplied by QDevil with a similar pass band. 

Each output line contains filters, isolators, and amplifiers. At the mixing chamber, we used an eccosorb filter, with part number matching that on the input line and the same K\&L filter as the input line. Two isolators were placed in series, both from QuinStar Technology (QCI-075900XM00). At the 4~K stage, a high electron mobility transistor (HEMT) amplifier was used (Low Noise Factory LNF-LNC4\_8F). Superconducting NbTi wire was used between the isolators and the HEMT to reduce signal attenuation. Additional filters were sometimes placed in the output line, with pass bands which contained all resonators that were being measured.

Several devices used a traveling wave parametric amplifier (TWPA) sourced from the MIT Lincoln Laboratory. The TWPA was placed in a separate magnetic shield at the mixing chamber and placed in the signal path immediately after the second isolator. The TWPA was pumped using a separate input line.

Outside of the fridge, we used additional amplifiers on the output line. We used amplifiers from Mini-Circuits (PNs: CMA-83LN+ and ZVE-3W-183+), RF-Lambda (PN: RLNA02G08G30), and Miteq (PN: AFS4-00101200-18-10P-4). The configuration of these amplifiers varied between experiments, but approximately 50 dB of gain was used in all cases.

All measurements were conducted with a vector network analyzer from Keysight (PNA-X Network Analyzer N5241A). For most experiments, the measurement parameters were: span of 5 times the high power resonator linewidth, 201 points across the frequency axis, IF bandwidth of 30 Hz, and an integration time per resonator varying from 1 minute (high power) to three hours (low power). Integration times were adjusted for each resonator chip, and measurement parameters differed slightly for early experiments.

\subsection{Resonator spectroscopy}

\begin{figure}[htbp]
    \centering
    \includegraphics[width=3.375in]{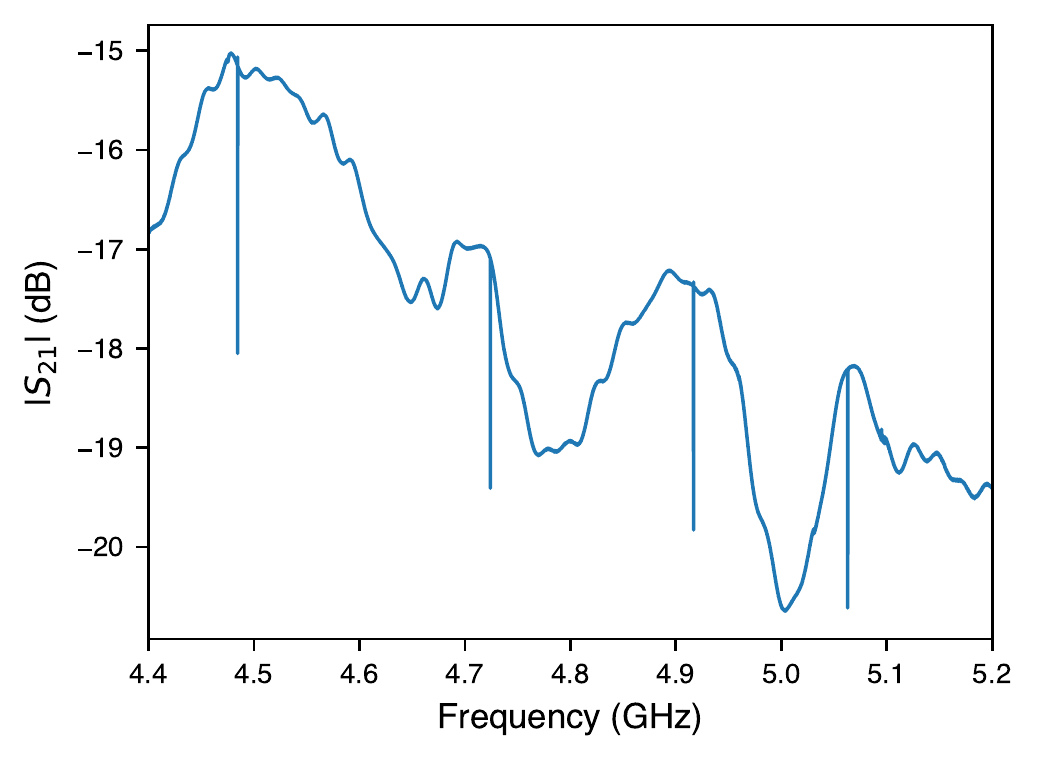}
    \caption{Wide frequency transmission sweep of a chip with four resonators coupled to a single feedline. The four sharp dips correspond to the location of the four resonators.}
    \label{fig:sup_locate_resonators}
\end{figure}

Resonators are easily located in frequency space due to their high quality factor relative to all other features. Figure 
\ref{fig:sup_locate_resonators} shows a wide frequency sweep of a chip with four resonators coupled to a feedline. The wide frequency ripples may be caused by standing waves or reflections from connections on our measurement setup; however, given that the width of these ripples are on the order of 10~MHz and the width of the resonators are on the order of 1~kHz, we ignore these ripples and assume that a flat background exists when measuring each resonator.

\subsection{Measuring $Q_{\textrm{int}}$} \label{sec:QintModel}

We used the following model to fit each resonator trace, such as the one shown in Figure 1c \cite{geerlings_improving_2013}:

\begin{equation} \label{eq:QintEquation}
    |S_{21}(\omega_{\textrm{probe}})| = \left|1-\frac{ \frac{Q_{tot}}{Q_c}-\frac{2iQ_{\textrm{tot}}\alpha}{\omega_0}}{1+2iQ_{\textrm{tot}}\frac{\omega_{\textrm{probe}}-\omega_0}{\omega_0}}\right| + S_{21,\textrm{baseline}},
\end{equation}

where $|S_{21}|$ is the magnitude of the transmission through the feedline, $Q_c$ is the coupling quality factor, $Q_{\textrm{tot}}$ is the total quality factor ($Q_{\textrm{tot}}^{-1} = Q_{\textrm{Int}}^{-1} + Q_c^{-1}$), $\alpha$ is the asymmetry of the resonator, $\omega_0$ is the center angular frequency of the resonator, $\omega_\textrm{probe}$ is the angular frequency of the probe tone, and $|S_{21,\textrm{baseline}}|$ is the transmission through the feedline when no resonator is present. We have assumed that $|S_{21,\textrm{baseline}}|$ is a constant, which is approximately correct for resonators with a small linewidth. The derivation of this model is given in the appendix of \cite{geerlings_improving_2013} with a minimal assumption set.

The coupling quality factor, $Q_c$, parameterizes the loss from the resonator to the feedline. In order to characterize our material losses, we must be able to separately determine $Q_c$ and $Q_{\textrm{int}}$ across an entire temperature and power sweep. The value of $Q_c$ is determined by the capacitive or inductive coupling between each resonator and the feedline, and therefore we expect $Q_c$ to be independent of power and temperature.

In our analysis, each resonator $|S_{21}|$ trace is fit independently. To show that we can separately extract $Q_c$ and $Q_\textrm{int}$ from the same $|S_{21}|$ trace, we examine the fitted values of $Q_c$ for each temperature sweep. We find that our fitted values of $Q_c$ are constant across power and temperature, and so we conclude that we have extracted an accurate value of $Q_c$, and therefore of $Q_\textrm{int}$. An example plot of $Q_c$ versus power and temperature, corresponding to the same sweep shown in Figure 2a, is shown below in Figure \ref{fig:sup_Qc}.

\begin{figure}[htbp]
    \centering
    \includegraphics[width=3.5in]{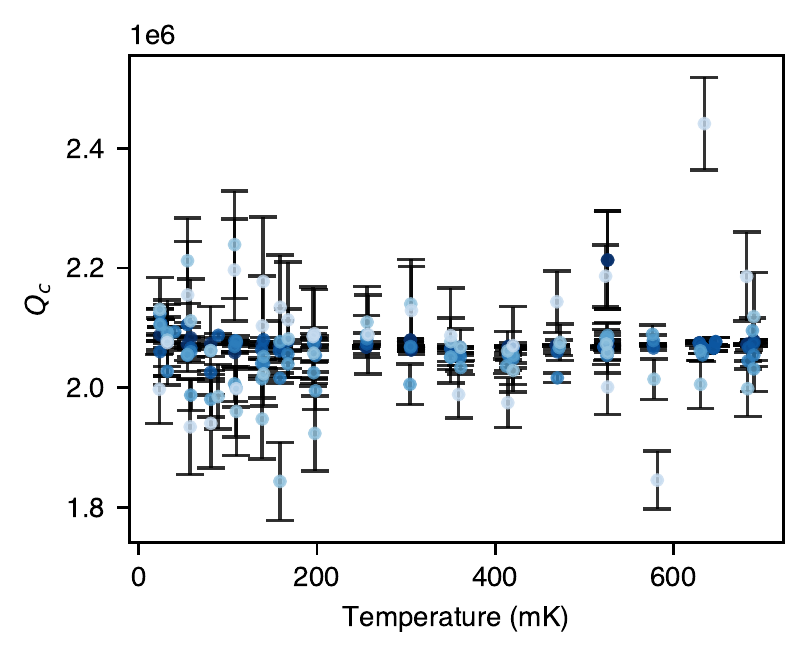}
    \caption{Fitted $Q_c$ parameters for a power and temperature sweep. This dataset corresponds to the same power and temperature sweep shown in Figure 2a of the main text. Colors correspond to power applied at the input, with the highest power being the darkest shade and the lowest power being the lightest shade. All powers are spaced 10~dB apart.}
    \label{fig:sup_Qc}
\end{figure} 

\subsection{Nonlinear behavior at high microwave power}

\begin{figure}[htbp]
    \centering
    \includegraphics[width=4.5in]{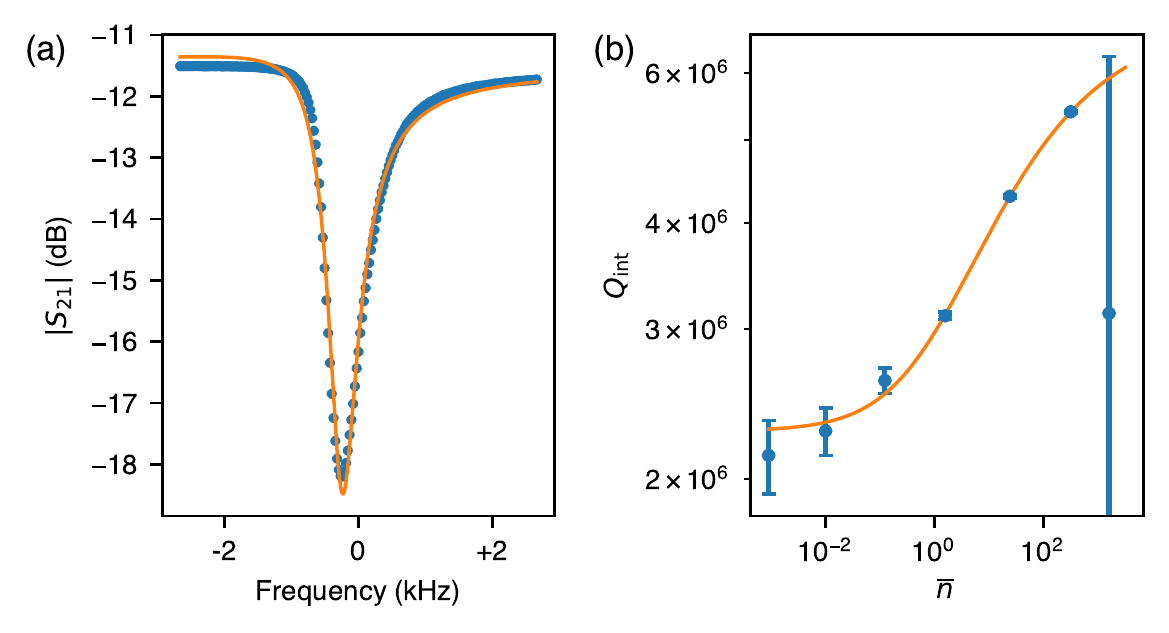}
    \caption{(a) An example of high-power resonator transmission trace with observable non-linearity. The best fit line to the data using Equation \ref{eq:QintEquation} is shown in orange. This trace was excluded from our analysis. (b) Power sweep showing $Q_\textrm{int}$ for the non-linear trace in (a) at the maximum $\overline{n}$. The orange line is a fit to the data with the non-linear datapoint excluded.}
    \label{fig:sup_nonLinear}
\end{figure}

When measuring the highest powers, we occasionally were unable to fit a resonator trace (Figure \ref{fig:sup_nonLinear}), which we attribute to nonlinear behavior of the resonator. Potential sources of this non-linearity are the saturation of an amplifier, or an effect of the superconducting state such as the non-linear kinetic inductance of Cooper pairs \cite{parmenter_nonlinear_1962, gittleman_nonlinear_1965}.

As we are most concerned with the behavior of our devices at low power, we excluded traces showing the nonlinear behavior from our analysis.

\subsection{Model for Q$_{\textrm{TLS}}$}
It can be shown that the loss induced by an ensemble of TLSs coupled to an electromagnetic mode takes the form \cite{gao_jiansong_physics_2008}:
\begin{equation}
    \frac{1}{Q_{\textrm{TLS}}(\overline{n}, T)} = \frac{1}{Q_\textrm{TLS,0}} \left( \frac{\tanh \left(\frac{\hbar \omega}{2 k_B T} \right)}{\sqrt{1 + \frac{\overline{n}}{n_c} T_1 T_2}} \right)
\end{equation}
where $\overline{n}$ is the average photon number in the mode, $T$ is the temperature of the mode-ensemble system, $\omega$ is the frequency of the mode, $n_c$ is the critical photon number of the ensemble, and $T_1$ and $T_2$ are the average relaxation and decoherence times of the ensemble. In order to obtain the model we use to fit the TLS component of our $Q_{\textrm{int}}$ data, we make a few substitutions. First, the average $T_1$ of the ensemble can be shown to follow a thermal distribution \cite{gao_jiansong_physics_2008}: 
\begin{equation}
    T_1 \propto \tanh \left( \frac{\hbar \omega}{2 k_B T} \right)
\end{equation}
Second, TLS-TLS interactions can be modeled as state changes in one TLS causing dephasing in neighboring TLS's. As the temperature is reduced, thermal fluctuations in the states of the TLSs in the ensemble will reduce as more and more members of the ensemble occupy the ground state. We therefore expect an inverse relationship between the TLS coherence time $T_2$ and temperature, which we model as \cite{burnett_evidence_2014}:
\begin{equation}
    \frac{1}{T_2} \propto k_B T^{\beta_1},
\end{equation}
where $\beta_1$ is an empirical parameter.

Finally, different mode shapes will overlap with and saturate the ensemble differently as they are populated with increasing numbers of photons, and we account for this by introducing another empirical fit parameter, $\beta_2$\cite{wang_improving_2009}:
\begin{equation}
    \overline{n} \rightarrow \overline{n}^{\beta_2}.
\end{equation}
Putting all of these substitutions together gives our TLS loss model:
\begin{equation}
    \frac{1}{Q_{\textrm{TLS}}(\overline{n}, T)} = \frac{1}{Q_\textrm{TLS,0}} \left( \frac{\tanh \left(\frac{\hbar \omega}{2 k_B T} \right)}{\sqrt{1 + \left( \frac{n^{\beta_2}}{D T^{\beta_1}} \right) \tanh \left(\frac{\hbar \omega}{2 k_B T} \right) }} \right)
\end{equation}
The model we use to fit quasiparticle losses has been discussed in other works \cite{zmuidzinas_superconducting_2012}.

\subsection{Model for frequency shift}
The model we use to fit the frequency shift is:
\begin{equation}
    \frac{\delta f(T)}{f_0} = \left( \frac{\delta f(T)}{f_0} \right)_\textrm{TLS} + \left( \frac{\delta f(T)}{f_0} \right)_\textrm{QP},
\end{equation}
where:
\begin{equation} \label{eq:fShift_TLS}
        \left( \frac{\delta f (T)}{f_0} \right)_\textrm{TLS} = \frac{1}{\pi Q_\textrm{TLS,0}} \textrm{Re} \left[ \Psi \left( \frac{1}{2} + i \frac{\hbar \omega}{2 \pi k_B T} \right) - \ln \left( \frac{\hbar \omega}{2 \pi k_B T} \right) \right] 
\end{equation}
is the TLS contribution to the frequency shift and:
\begin{equation}\label{eq:fShift_QP}
    \left( \frac{\delta f (T)}{f_0} \right)_\textrm{QP} = -\frac{\alpha}{2} \left( 1 - \sin \left( \phi (T, \omega) \right) \sqrt{ \frac{\sigma_1 (T, \omega)^2 + \sigma_2 (T, \omega)^2}{\sigma_1 (0, \omega)^2 + \sigma_2 (0, \omega)^2} } \right)
\end{equation}
is the quasiparticle contribution to the frequency shift. In Equation \ref{eq:fShift_TLS} and \ref{eq:fShift_QP}, $\Psi$ is the complex digamma function; $\sigma_1$ and $\sigma_2$ are the real and imaginary parts of the complex conductivity; $\phi$ is the phase between the real and imaginary parts of the complex conductivity; and $\alpha$ is the kinetic inductance fraction. The derivation of the TLS contribution can be found in, for example, \cite{gao_jiansong_physics_2008}. The expression for the quasiparticle contribution is based on \cite{gao_jiansong_physics_2008} but is not explicitly stated, so we derive it in detail below. 
\par The frequency shift from quasiparticles is defined as:
\begin{equation}
    \frac{\delta f(T)}{f_0} = - \frac{\alpha}{2} \left( \frac{X_S (T) - X_S (0)}{X_S (0)} \right),
\end{equation}
where $X_S$ is the imaginary part of the surface impedance of the superconductor, otherwise known as the reactance. In general the surface impedance has a cumbersome form, but in three superconducting material limits it takes the simpler form:
\begin{equation}
    Z_s (T) = A \sigma(T)^\gamma
\end{equation}
where $A$ is a constant prefactor, $\sigma(T)$ is the superconducting complex conductivity, and $\gamma$ is a parameter that takes a different value depending on which of the three limits the superconductor is in. As the quasiparticles of interest in our system have a thermal distribution, the complex conductivity takes the form:
\begin{equation}
    \sigma(T) = \sigma_1 (T) + i \sigma_2 (T)
\end{equation}
where:
\begin{equation}
    \frac{\sigma_1(T)}{\sigma_n} = \frac{4 \Delta_0}{\hbar \omega} e^{-\Delta_0/k_B T} \sinh\left( 
\frac{\hbar \omega}{2 k_B T} \right) K_0 \left( \frac{\hbar \omega}{2 k_B T} \right)
\end{equation}
and:
\begin{equation}
    \frac{\sigma_2(T)}{\sigma_n} = \frac{\pi \Delta_0}{\hbar \omega} \left[ 1 - \sqrt{\frac{2 \pi k_B T}{\Delta_0}} e^{-\Delta_0 / k_B T} - 2e^{-\Delta_0 / k_B T} e^{-\hbar \omega / 2 k_B T} I_0 \left( \frac{\hbar \omega}{2 k_B T} \right) \right]
\end{equation}
In the above equations, $\hbar$ is the reduced Planck constant, $k_B$ is the Boltzmann constant, $\omega$ is the resonator center angular frequency, $T$ is temperature, $\Delta_0 = 1.764 k_B T_c$ is the superconducting gap, $T_c$ is the superconducting critical temperature, $I_0$ is the zeroth order modified Bessel function of the first kind, $K_0$ is the zeroth order modified Bessel function of the second kind, and $\sigma_n$ is the normal-state conductivity of the superconductor just above $T_c$. The surface impedance can be rewritten in a more convenient form using these quantities:
\begin{equation}
    \begin{split}
        Z_s(T) & = A (\sigma_1 (T) + i \sigma_2 (T))^\gamma \\
        & = A (|\sigma (T)| e^{i \phi(T)})^\gamma \\
        & = A |\sigma (T)|^\gamma e^{i \gamma \phi(T)}) \\
        & = A |\sigma (T)|^\gamma (\cos(\gamma \phi (T)) + i \sin(\gamma \phi (T)))
    \end{split}
\end{equation}
where:
\begin{equation}
    |\sigma (T)| = \sqrt{\sigma_1(T)^2 + \sigma_2(T)^2}
\end{equation}
and:
\begin{equation}
    \phi(T) = \arctan \left( \frac{\sigma_2(T)}{\sigma_1(T)} \right).
\end{equation}
The reactance is then:
\begin{equation}
    \begin{split}
        X_S(T) & = \textrm{Im} [Z_S(T)] \\
        & = A |\sigma (T)|^\gamma (\sin(\gamma \phi (T))).
    \end{split}
\end{equation}
The frequency shift can then be written in terms of the complex conductivities:
\begin{equation}
    \begin{split}
        \frac{\delta f(T)}{f_0} & = - \frac{\alpha}{2} \left( \frac{X_S (T) - X_S (0)}{X_S (0)} \right) \\
        & = - \frac{\alpha}{2} \left( 1 - \frac{\sin (\gamma \phi(T))}{\sin (\gamma \pi / 2)} \sqrt{\frac{|\sigma (T)|}{|\sigma (0)|}} \right) \\
    \end{split}
\end{equation}
There are three possible values of $\gamma$ depending on the electron mean free path $(\ell)$, coherence length $(\xi_0)$, film thickness $(d)$, and London penetration depth $(\lambda_{\textrm{LO}})$ of the superconductor \cite{gao_jiansong_physics_2008}:
\begin{equation}
    \begin{split}
        \gamma & = -1/3, \, \textrm{thick film, extreme anomalous limit} \,\, (\xi_0 \gg \lambda_{\textrm{LO}} \,\, \textrm{AND} \,\, \ell \gg \lambda_{\textrm{LO}}), \\
        \gamma & = -1/2, \,\, \textrm{thick film, dirty limit} \,\, (\ell \ll \xi_0 \,\, \textrm{AND} \,\, \ell \ll \lambda_{\textrm{LO}}), \\
        \gamma & = -1, \textrm{thin film, dirty limit} \,\, (d \sim \ell \ll \xi_0 \,\, \textrm{AND} \,\, d \sim \ell \ll \lambda_{\textrm{LO}}). \\
    \end{split}
\end{equation}
However, we did not directly measure the relevant parameters to determine whether or not we were in any of these three regimes, so instead we fit our data to all three to see if the assumed regime made a difference to the outcome of the fit. The results of this analysis are shown below, where the consistency between the superconducting critical temperature $T_c$ estimated by our $Q_\textrm{int}$ fits and our frequency shift fits is plotted for all three values of $\gamma$. 

\begin{figure}[htbp]
    \centering
    \includegraphics{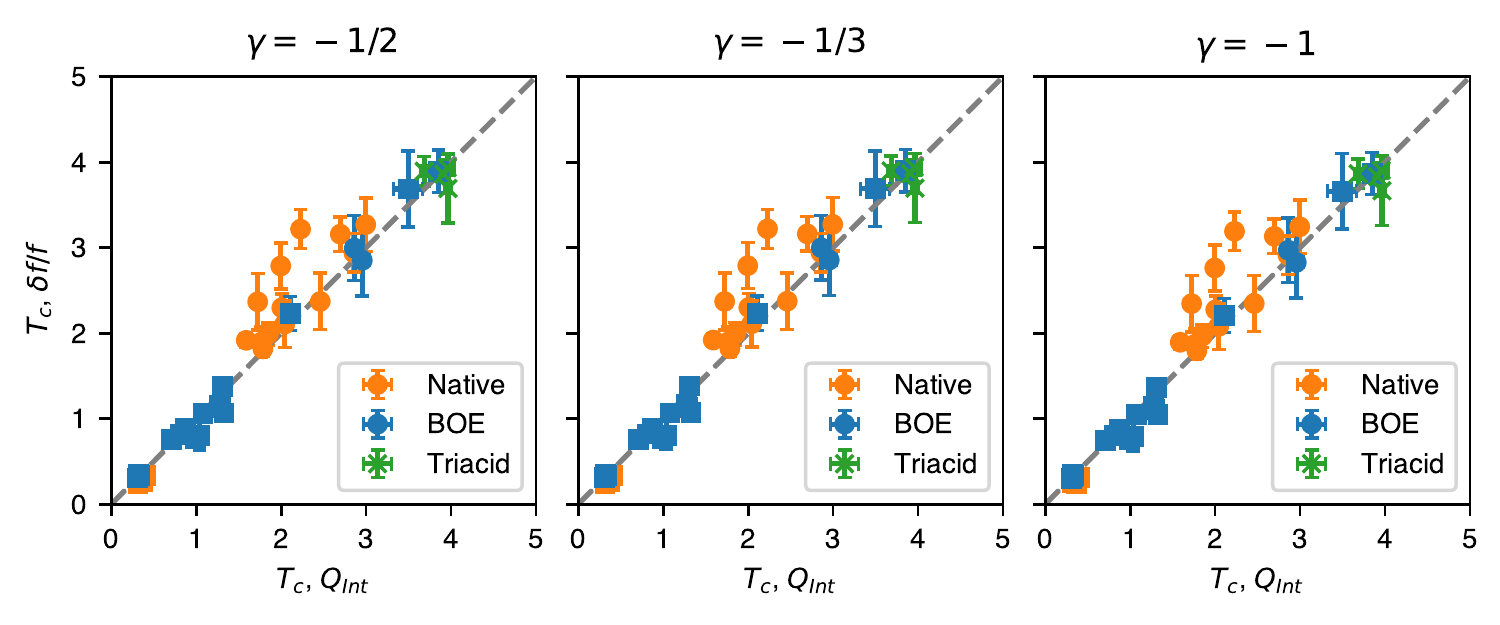}
    \caption{Plots of $T_c$ estimates provided by the frequency shift and $Q_\textrm{int}$ fitting methods for the three different quasiparticle frequency shift regimes.}
    \label{fig:tcConsistency}
\end{figure}

As can be seen, the assumed regime makes no difference to the fit outcome, so we choose to work in the thin film local limit ($\gamma = -1$) for our quasiparticle frequency shift fits: 
\begin{equation}
    \left( \frac{\delta f (T)}{f_0} \right)_\textrm{QP} = -\frac{\alpha}{2} \left( 1 - \sin \left( \phi (T, \omega) \right) \sqrt{ \frac{\sigma_1 (T, \omega)^2 + \sigma_2 (T, \omega)^2}{\sigma_1 (0, \omega)^2 + \sigma_2 (0, \omega)^2} } \right).
\end{equation}

\section{Resonator design}
\subsection{CPW resonators}
Our CPW resonators are quarter-wave resonators constructed by shorting one end of a transmission line. Our design sets the characteristic impedance ($Z_0$) of the resonators to 50 $\Omega$, which dictates a relationship between the centerpin width and the gap width \cite{simons_coplanar_2008}. This means that if the distance between the center pin and ground plane (pitch) of the resonator is specified, the centerpin width is fully constrained. The resonators are designed to have resonance frequencies between 6 and 8 GHz, where the resonance frequency is dictated by:
\begin{equation}
    f_0 = \frac{v}{4 l \sqrt{\epsilon_{\textrm{eff}}}},
\end{equation}
where $\epsilon_{\textrm{eff}}$ is the effective dielectric constant defined in \cite{simons_coplanar_2008}, $l$ is the length of the resonator, and $v$ is the speed of electric field propagation down the transmission line. We generally assume $v = c$, where $c$ is the speed of light in vacuum.
\par When designing both the LE and CPW resonators, we aim to have the coupling rate of the resonator to the feedline ($1/Q_c$, sometimes written as $1/Q_\textrm{ext}$) be equal to the expected internal loss rate of the resonator ($Q_\textrm{int}$). If $Q_c$ is too small, the measurement is not sensitive to changes in $Q_{int}$; and if $Q_c$ is too large, the photon lifetime in the resonator is short and signal-to-noise (SNR) decreases. Our CPW resonators are capacitively coupled to the feedline, and we compute this coupling using an equation from \cite{palacios-laloy_superconducting_2010}:
\begin{equation}
    C_c = \sqrt{\frac{\pi}{4 Q_c}} \frac{1}{2 \pi f_0 Z_0},
\end{equation}
where $C_c$ is the capacitance between the centerpin and the feedline. We compute this capacitance using finite element analysis (ANSYS Maxwell 3D), and generally find good agreement between predicted and measured resonance frequencies and external loss rates.

\subsection{LE resonators}
The LE resonators consist of a meander inductor in series with a dipole capacitor, with a resonance frequency given by:
\begin{equation}
    f_0 = \frac{1}{\sqrt{L C}}
\end{equation}
and an impedance given by:
\begin{equation}
    Z_0 = \sqrt{\frac{L}{C}},
\end{equation}
where $L$ and $C$ are the inductance of the inductor and the capacitance of the capacitor. We account for stray capacitance across the inductor by modeling a stray capacitor in parallel with the lumped inductor. Therefore the total capacitance in the resonator is the sum of the lumped capacitance $C_L$ and the stray capacitance $C_S$, $C_\textrm{tot} = C_L + C_S$. The resonance frequency and impedance are then:
\begin{equation} \label{eq:f0_LE}
    f_0 = \frac{1}{\sqrt{(C_L + C_S) L}}
\end{equation}
and:
\begin{equation}
    Z_0 = \sqrt{\frac{L}{C_L + C_S}}.
\end{equation}
For a given resonator design, we compute these three unknowns using three separate simulations. The first is a capacitance simulation in Ansys Maxwell 3D of only the dipole capacitor pads. We take the modeled capacitance to be equal to the lumped capacitance $C_L$. The second and third simulations are HFSS eigenmode simulations of the meander and the full resonator. The resonance frequency of the meander can be written as:
\begin{equation}
    f_{0, \textrm{meander}} = \frac{1}{\sqrt{L C_S}}
\end{equation}
and the resonance frequency of the full resonator is given by Equation \ref{eq:f0_LE}.
With $C_L$, $f_{0, \textrm{meander}}$ and $f_{0, \textrm{resonator}}$ calculated from the three simulations, the remaining unknowns ($C_S$ and $L$) and the fundamental resonator parameters ($f_0$ and $Z_0$) can be computed. 
\par The external coupling rate of the LE resonators was determined empirically by cooling down an initial design, and then adjusting the distance from the feedline to better match the external coupling rate to the internal loss rate. The distance from the feedline was adjusted by assuming the coupling would fall off proportional to $1/r^3$, as the coupling is inductive. After this initial cooldown, finer adjustments were made for subsequent designs, but in general the external coupling rate of the LE resonators matched the internal loss rate, which is the aforementioned condition for optimizing both SNR and sensitivity to changes in $Q_{\textrm{int}}$. 

\section{SPR calculations}
For both the CPW and LE resonators, the SPRs reported in the main text are computed by simulating the electric field energy stored in $3$ nm thick dielectric interface layers with dielectric constants of $\epsilon=10$. For the CPW resonators, the simulation is done using a single cross section of the centerpin and ground plane, and for the LE resonators the simulation is done using a single cross section of the dipole capacitor pads. We use DC finite element simulations (Ansys Maxwell) for both kinds of single cross section simulations.
\par The single cross section approximation is appropriate for the CPW resonators because their geometry is a single cross section extruded along a path. However, the degree to which the single cross section simulation is a good approximation for the LE resonators is not as obvious, as the LE resonators have nontrivial structure in the direction normal to the cross section plane. To check that the single cross section simulations accurately estimate the SPR's of the LE resonators, we separately compute the SPRs for a handful of LE resonators using the method outlined in \cite{wang_surface_2015}, which involves an eigenmode simulation supplemented with a DC cross section simulation of the metal edges. We find that the single cross section and 2D sheets methods agree to within 15\%, indicating that the single dipole capacitor cross section simulation is a suitable approximation for the full LE resonator SPR. This also implies that the meander inductor does not contribute significantly to the total SPR of the LE resonators.

\section{Bulk losses} \label{sec:bulk}

In the main text we describe how we extract loss independent of SPR by fitting $Q_\textrm{TLS,0}$ versus SPR, and we extract a low power bulk loss tangent an order of magnitude larger than that measured in \cite{read_precision_2022}.

One hypothesis for this difference is that the ``bulk" loss to which our measurements are sensitive is not the same as the volumetric average bulk loss measured by \cite{read_precision_2022}. In our experiments, the device with the lowest SPR is an LE device with 65~$\mu$m spacing between capacitor pads. Our experiment therefore cannot distinguish between ``bulk" and depths below the surface comparable to this spacing. In \cite{read_precision_2022}, by contrast, the experiment probes the loss tangent averaged over the bulk of a 440~$\mu$m thick HEMEX sample. We hypothesize that a near surface layer hosts a higher concentration of defects that give rise to TLS behavior. Since our measured bulk loss tangent is an order of magnitude higher than measured in \cite{read_precision_2022}, in order to reconcile the two measurements this highly damaged layer would need to be around ten times thinner than the bulk substrate measured in \cite{read_precision_2022}, around 50~$\mu$m.

These hypothesized extended defects could be caused by polishing, damage from etching, or other fabrication induced damage. Direct materials characterization of the polished sapphire could elucidate potential microscopic sources of TLS associated with this damage.

\section{Surface characterization after chemical processing} \label{sec:XPS}
We used X-ray photoelectron spectroscopy (XPS) to characterize the surface of our tantalum films before and after surface processing. We started with samples that had hard baked photoresist applied and stripped off in solvent following the procedures outlined in Section \ref{sec:fab}. We scanned a sample before any further chemical processing, after a piranha treatment (``native" surface), and after both a piranha and a 20 minute BOE treatment (``BOE" surface), as well as a separate sample after triacid treatment (``triacid" surface). To reduce the amount of adventitious carbon accumulated on the samples after chemical cleaning, we attempted to keep the length of time between chemical treatments and XPS measurements low. We took measurements within thirty minutes of the piranha treatment and BOE treatment, and took measurements four hours after triacid treatment. All XPS measurements were taken on a ThermoFisher K-Alpha XPS Spectrometer with an aluminum K$\alpha$ X-ray source.

We took a broad survey scan on each sample and observed Ta, O, and C peaks on all samples, and a Na1s peak on the untreated sample only. We took fine scans of the Ta4f, O1s, and C1s peaks for all samples with a binding energy step size of 0.1~eV and a dwell time of 50~ms. We subtracted a Shirley background from the Ta4f and O1s peaks \cite{engelhard_introductory_2020}, and a linear background from the C1s peaks. To account for potentially different X-ray flux between different measurements, we normalized all intensity data to the total intensity of the Ta4f spectrum for each sample. In addition, we calibrate the binding energy scale by setting the lowest binding energy Ta4f peak to 21.2~eV.

\begin{figure}[htbp]
    \centering
    \includegraphics[width=6.75in]{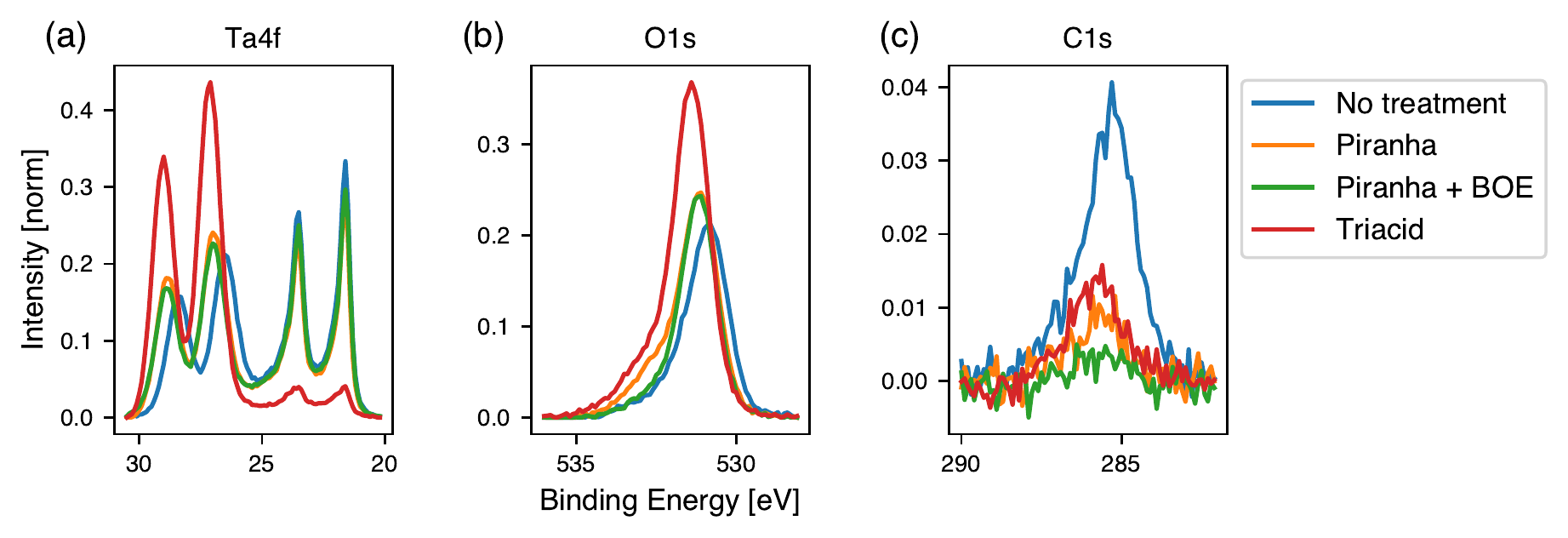}
    \caption{XPS data of the surface of tantalum films after various surface treatments. Scans are of the Ta4f (a), O1s (b), and C1s (c) spectra. The ``No treatment" data were taken after photoresist was stripped from a sample. ``Piranha", ``Piranha + BOE", and ``Triacid" correspond to the ``Native", ``BOE", and ``Triacid" surface conditions, respectively. Ta4f and O1s data have a Shirley background subtracted \cite{engelhard_introductory_2020}, and all C1s data have a linear background subtracted. All data are normalized to the total Ta4f intensity measured on the sample.}
    \label{fig:sup_XPS}
\end{figure}

In the Ta4f spectrum, we can resolve two pairs of two peaks. We attribute the symmetric pair of peaks between 26~eV and 30~eV to the dominant Ta$^{5+}$ oxidation state and the asymmetric pair of peaks between 21~eV and 24~eV to the tantalum metal \cite{himpsel_core-level_1984}. Each state generates two peaks due to the strong spin-orbit coupling in tantalum \cite{Moulder:1992}. The relative intensity of the Ta$^{5+}$ peaks is smallest for the untreated sample, increases slightly after a piranha treatment, decreases slightly after a BOE treatment, and is largest after a triacid treatment (Figure \ref{fig:sup_XPS}(a), qualitatively matching what is described in \cite{mclellan_chemical_2023}).

For the C1s peak, the intensity is maximized for the untreated sample, and is significantly reduced by each acid treatment. Performing a BOE treatment after piranha treatment reduces the C1s intensity over that of just piranha. The measurement on our triacid treated sample shows a strongest C1s signal out of the three acid treated measurements (Figure \ref{fig:sup_XPS}(c)).
 
The relative intensity of the Ta$^{5+}$ doublet and the intensity of the O1s peak both indicate that the oxide thickness grows slightly after piranha treatment, is etched slightly after the BOE treatment, and is grown significantly after triacid treatment. In \cite{mclellan_chemical_2023}, we measure that the BOE treatment reduces the oxide thickness by 20\% and the triacid treatment grows the oxide thickness by over a factor of two.

The sources of carbon in our system are adventitious carbon and photoresist residue. Therefore, the intensity change of the C1s peak is related to removal of fabrication residue, but can be complicated by the duration of air exposure, which leads to adventitious carbon accumulation. Our data shows that piranha treatment is effective at removing carbon from the surface, but performing BOE in addition to piranha can remove more carbon than piranha alone. We attribute this further reduction to carbon being removed from the surface of the tantalum oxide as it is etched away. We expect the triacid treatment to be extremely effective at cleaning the surface \cite{sangtawesin_origins_2019}, however, the measurement of the triacid treated sample does not show as much reduction in the C1s signal. We attribute the larger triacid signal to the increased length of time between cleaning and measurement, which would allow more adventitious carbon to deposit on the surface.

\section{Model for surface losses} \label{sec:loss_model}
In the limit that surface losses dominate, dielectric loss in superconducting resonators can be expressed as:
\begin{equation}
    \frac{1}{Q_{\textrm{TLS},0}} = p_{\textrm{MS}} \tan \delta_{\textrm{MS}} + p_{\textrm{MA}} \tan \delta_{\textrm{MA}} + p_{\textrm{SA}} \tan \delta_{\textrm{SA}}
\end{equation}
where $\tan \delta_i$ is the loss tangent of interface $i$; and MS, MA, and SA are the metal-substrate, metal-air, and substrate-air interfaces, respectively. The above expression can be rearranged as follows:
\begin{equation}
    \begin{split}
        \frac{1}{Q_{\textrm{TLS},0}} & = p_{\textrm{MS}} \tan \delta_{\textrm{MS}} + p_{\textrm{MA}} \tan \delta_{\textrm{MA}} + p_{\textrm{SA}} \tan \delta_{\textrm{SA}} \\
        & = p_{\textrm{MS}} \left( \tan \delta_{\textrm{MS}} + \beta_{\textrm{MA}} \tan \delta_{\textrm{MA}} + \beta_{\textrm{SA}} \tan \delta_{\textrm{SA}} \right) \\
        & = p_{\textrm{MS}} \tan \delta
    \end{split}
\end{equation}
where $\beta_i = p_i / p_{\textrm{MS}}$, and $\tan \delta$ is the parameter we fit for. We can recast the above in terms of $p_\textrm{MA}$:
\begin{equation}
    \begin{split}
        \frac{1}{Q_{\textrm{TLS},0}} & = p_{\textrm{MS}} \tan \delta \\
        & = p_{\textrm{MA}} \left( \frac{p_{\textrm{MS}}}{p_{\textrm{MA}}} \right) \tan \delta \\
        & = p_{\textrm{MA}} \alpha_{\textrm{MS}} \tan \delta \\
    \end{split}
\end{equation}
where $\alpha_i = p_i / p_{\textrm{MA}}$.

\begin{figure}
    \centering
    \includegraphics[width=\columnwidth]{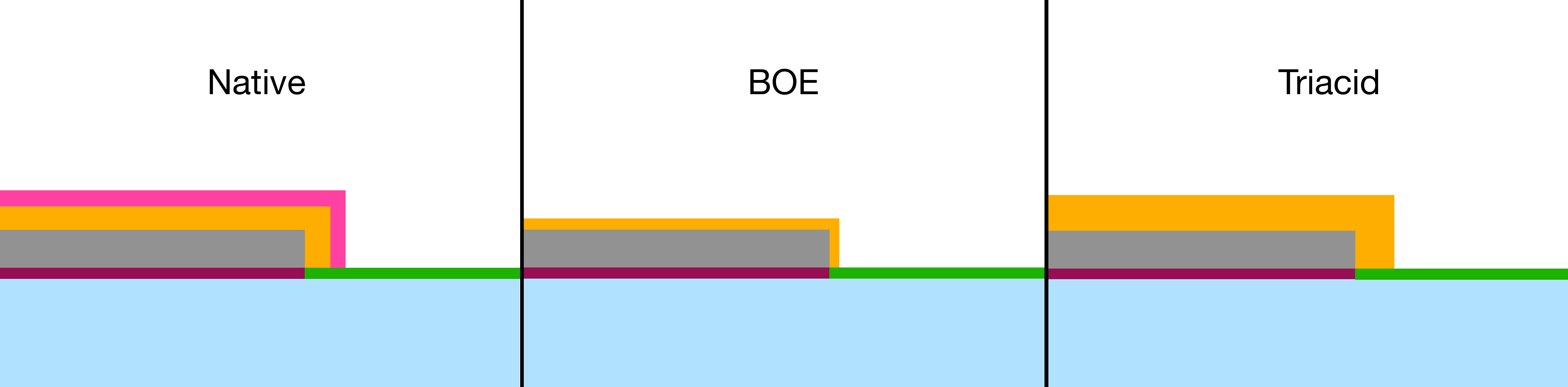}
    \caption{A model for hydrocarbon losses in which BOE and triacid treatments remove residual hydrocarbons left over from photoresist. The native samples suffer from losses from both the native oxide and the hydrocarbons on the MA interface. In the above cartoon, the pink layer is hydrocarbons, and the orange, green, and purple layers are the MA, SA and MS interfaces, respectively. The ``BOE" diagram corresponds both to the BOE and long BOE treatments, with the difference being the thickness of the oxide layer. This model assumes that piranha cleaning is effective at removing hydrocarbons on the sapphire, but not on the oxide surface.}
    \label{fig:nativeMAonly}
\end{figure}

We now consider a model in which the BOE, long BOE, and triacid samples have a source of loss on the MA interface that scales linearly with the oxide thickness, and the native samples suffer from both this oxide-thickness dependent loss and an additional source of loss on the MA interface which we hypothesize is due to fabrication related hydrocarbons (Figure \ref{fig:nativeMAonly}). We can recast losses in terms of the true oxide thickness and the hydrocarbon related loss by:
\begin{equation}
    \begin{split}
        \left(\frac{1}{Q_{\textrm{TLS},0}}\right)_i & = p_{\textrm{MA}} \alpha_{\textrm{MS}} \tan \delta_i \\
        & = p_{\textrm{MA}} \left( \frac{t_i}{t_0} \right) \left( \frac{t_0}{t_i} \right) \alpha_{\textrm{MS}} \tan \delta_i \\
        & = p_{\textrm{MA,i}} \left( \frac{t_0}{t_i} \right) \alpha_{\textrm{MS}} \tan \delta_i \\
        & = p_{\textrm{MA,i}} \left( \frac{t_0}{t_i} \right) \left( \tan \delta_{\textrm{MA}} + \alpha_{\textrm{SA}} \tan \delta_{\textrm{SA}} + \alpha_{\textrm{MS}} \tan \delta_{\textrm{MS}}  + \gamma_i \tan \delta_{\textrm{HC}} \right) \\
        & = p_{\textrm{MA,i}}  \left( \tan \delta_{\textrm{MA,0}} + \left( \frac{t_0}{t_i} \right) (\alpha_{\textrm{SA}} \tan \delta_{\textrm{SA}} + \alpha_{\textrm{MS}} \tan \delta_{\textrm{MS}} + \gamma_i \tan \delta_{\textrm{HC}}) \right) \\
    \end{split}
\end{equation}
where $t_0 = 3$ nm is the standard assumed oxide thickness, $t_i$ is the measured oxide thickness for the $i^{\textrm{th}}$ surface processing technique, $p_{\textrm{MA, i}}$ is the true MA surface participation of the $i^{\textrm{th}}$ surface processing technique (up to a factor of the assumed oxide dielectric constant, $\epsilon=10$), and $\gamma_i \in \{0,1\}$ determines if hydrocarbon loss is considered for the $i^{\textrm{th}}$ surface processing technique ($\gamma_\textrm{native} = 1$ and $0$ otherwise). Equating the third and fifth lines of the above gives:
\begin{equation} \label{eq:surf_loss_model}
    \alpha_{\textrm{MS}} \tan \delta_i = \left( \frac{t_i}{t_0} \right) \tan \delta_{\textrm{MA,0}} + \alpha_{\textrm{SA}} \tan \delta_{\textrm{SA}} + \alpha_{\textrm{MS}} \tan \delta_{\textrm{MS}} + \gamma_i \tan \delta_{\textrm{HC}}.
\end{equation}

By considering the native surface and any two of the BOE, long BOE, and triacid surface, we can solve the above equations for $\tan \delta_{\textrm{MA,0}}$, $\alpha_{\textrm{SA}} \tan \delta_{\textrm{SA}} + \alpha_{\textrm{MS}} \tan \delta_{\textrm{MS}}$, and $\tan \delta_{\textrm{HC}}$. 

Consider a set of surface treatments $\{ \textrm{Native}, a, b \}$, where $a$ and $b$ are any pair of BOE, long BOE, and triacid. The system of equations described by Equation \ref{eq:surf_loss_model} for this set of treatments is solved for $\tan \delta_{\textrm{MA,0}}$ and $\alpha_{\textrm{SA}} \tan \delta_{\textrm{SA}} + \alpha_{\textrm{MS}} \tan \delta_{\textrm{MS}}$ by:

\begin{equation}
    \begin{split}
        \tan \delta_{\textrm{MA,0}} = \left( \frac{t_0}{t_{\textrm{a}} - t_{\textrm{b}}} \right) \alpha_{\textrm{MS}} (\tan \delta_{\textrm{a}} - \tan \delta_{\textrm{b}})
    \end{split}
\end{equation}
and 
\begin{equation}
    \begin{split}
        \alpha_{\textrm{SA}} \tan \delta_{\textrm{SA}} + \alpha_{\textrm{MS}} \tan \delta_{\textrm{MS}} = \alpha_{\textrm{MS}} \left( \frac{t_{\textrm{a}} \tan \delta_{\textrm{b}}  - t_{\textrm{b}} \tan \delta_{\textrm{a}}}{t_{\textrm{a}} - t_{\textrm{b}}} \right). \\
    \end{split}
\end{equation}
Note that these solutions only involve the surface treatments $a$ and $b$. Solving for $\tan \delta_{\textrm{HC}}$:
\begin{equation}
    \tan \delta_{\textrm{HC}} = \alpha_{\textrm{MS}} \tan \delta_{\textrm{Native}} - \left( \frac{t_\textrm{Native}}{t_0} \right) \tan \delta_{\textrm{MA,0}} - (\alpha_{\textrm{SA}} \tan \delta_{\textrm{SA}} + \alpha_{\textrm{MS}} \tan \delta_{\textrm{MS}}).
\end{equation}

\begin{figure}
    \centering
    \includegraphics[width=\columnwidth]{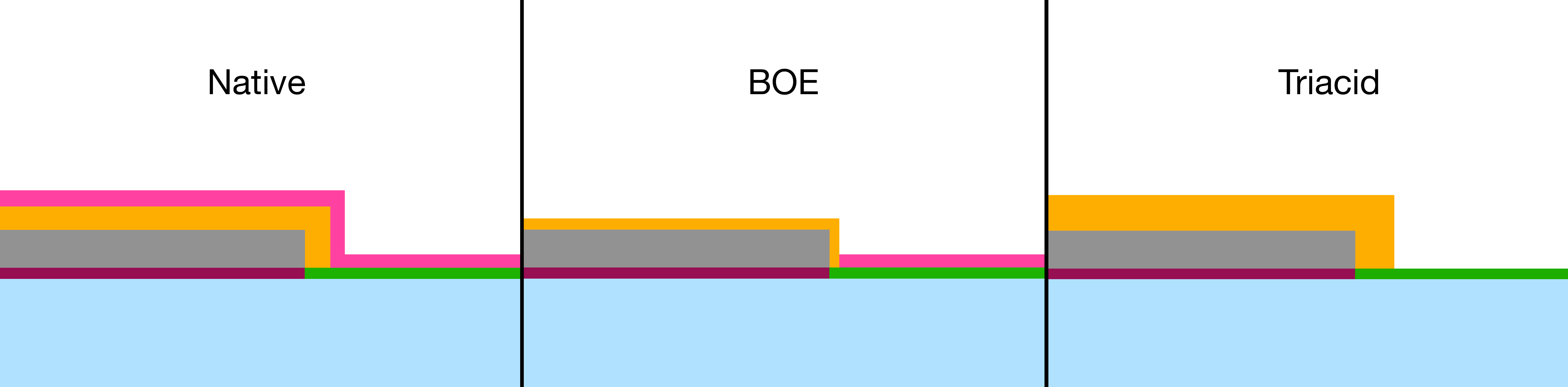}
    \caption{An example of a model excluded by our data. Since BOE does not etch sapphire and does not etch hydrocarbons, one possibility is that hydrocarbons reside on the MA and SA interfaces of the native samples and the SA interface of the BOE and long BOE. This model could be achieved if piranha cleaning was ineffective at removing hydrocarbons from the sapphire surface, while the triacid treatment is highly effective. We continue to assume that hydrocarbons are lifted off from tantalum with BOE etching of the oxide. The ``BOE" diagram corresponds both to the BOE and long BOE treatments, with the difference only being the thickness of the oxide layer. The parameter values we extract from this model given our data are unphysical.}
    \label{fig:excludedModel}
\end{figure}

In order to have a better basis of comparison to our extracted quantities, we rescale the quantities computed above to $p_{\textrm{MS}}$, which is the conventional metric by which surface-dependent losses are compared. For the hydrocarbon loss, we have:

\begin{equation}
    \begin{split}
        \frac{1}{Q_{\textrm{HC}}} & = p_{\textrm{MA}} \tan \delta_{\textrm{HC}} \\
        & = p_{\textrm{MS}} \beta_{\textrm{MA}} \tan \delta_{\textrm{HC}} \\
    \end{split}
\end{equation}
where $Q_{\textrm{HC}}$ is the inverse loss associated with the hydrocarbons and $\beta_\textrm{MA} = p_\textrm{MA} / p_{\textrm{MS}}$. 

For the SA and MS loss:
\begin{equation}
    \begin{split}
        \frac{1}{Q_\textrm{MS}} + \frac{1}{Q_\textrm{MA}} & = p_{\textrm{MA}} (\alpha_{\textrm{SA}} \tan \delta_{\textrm{SA}} + \alpha_{\textrm{MS}} \tan \delta_{\textrm{MS}}) \\
        & = p_{\textrm{MS}} \beta_{\textrm{MA}} (\alpha_{\textrm{SA}} \tan \delta_{\textrm{SA}} + \alpha_{\textrm{MS}} \tan \delta_{\textrm{MS}}) \\
    \end{split}
\end{equation}
where $Q_{\textrm{SA}}$ and $Q_{\textrm{MS}}$ are the inverse loss associated with the SA and MS interfaces.

\begin{figure}
    \centering
    \includegraphics[width=\columnwidth]{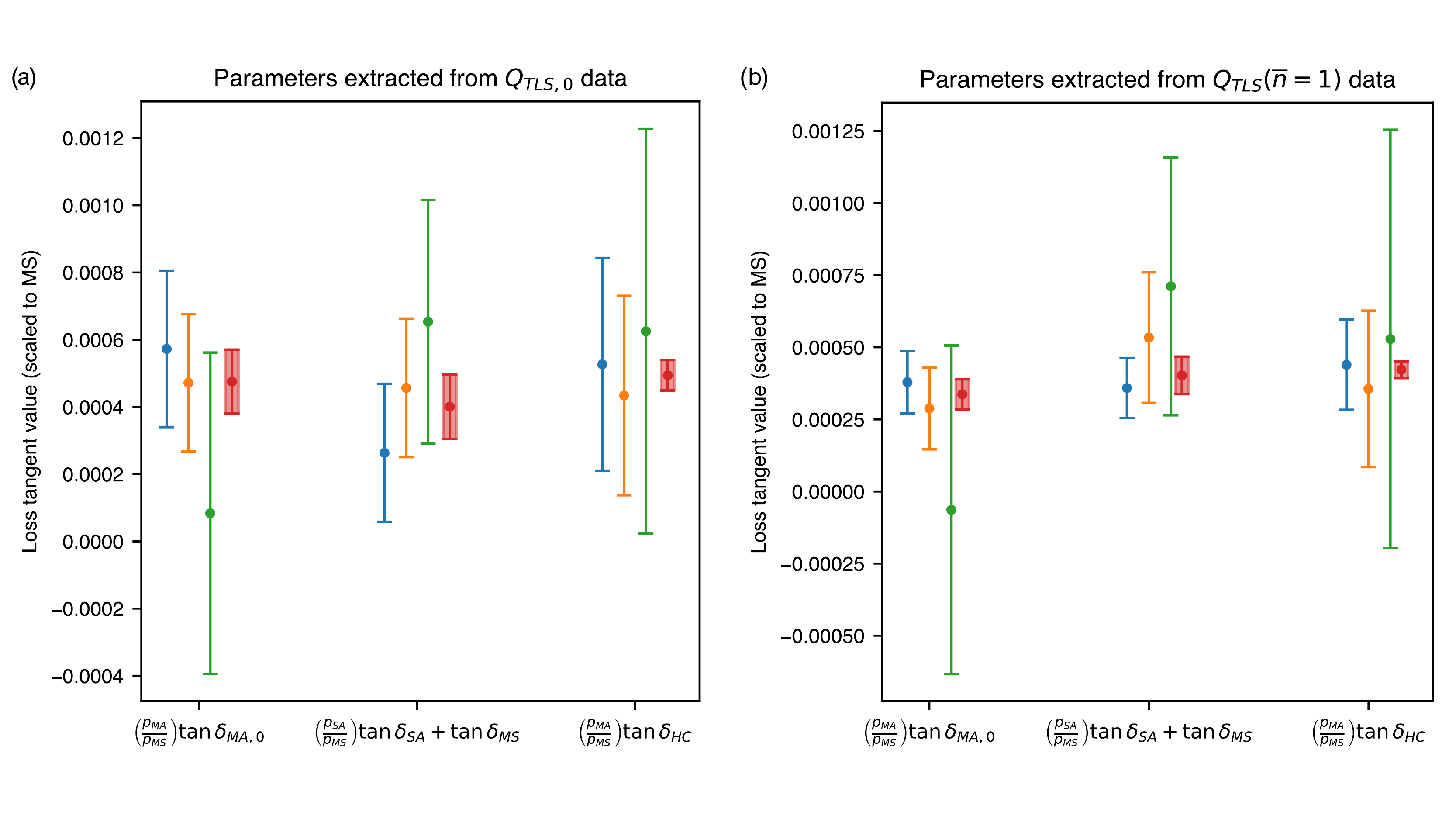}
    \caption{Model estimates provided by the three possible combinations of hydrocarbon-free species. Blue is BOE and triacid, orange is long BOE and triacid, and green is BOE and long BOE. The best fit value is shown in red, with a shaded box to distinguish it from the different extracted estimates. (a) Estimates from $Q_\textrm{TLS,0}$ data. (a) Estimates from $Q_\textrm{TLS}(\overline{n}=1)$ data.}
    \label{fig:triplets}
\end{figure}

Oxide thicknesses for the native, BOE, and triacid treatments are determined in \cite{mclellan_chemical_2023}, and we estimate the oxide thickness for the long BOE in Section \ref{sec:longBOEThick}. In all cases we consider the total oxide thickness to be a sum of the Ta$^{5{+}}$, Ta$^{3{+}}$, and Ta$^{1{+}}$ species. We compare the solutions for $\tan \delta_{\textrm{MA,0}}$, $\alpha_{\textrm{SA}} \tan \delta_{\textrm{SA}} + \alpha_{\textrm{MS}} \tan \delta_{\textrm{MS}}$, and $\tan \delta_{\textrm{HC}}$ for different choices of three surface treatments in Figure \ref{fig:triplets}, and find that the solutions agree to within uncertainties. For each parameter, we fit the best single value to the three values reported by the three possible sets of surface treatments, and report these fitted values in the main text.

Other assumptions about the configuration of hydrocarbons after the three surface treatments can be made, but we find that our data exclude certain configurations of hydrocarbons. For example, if we consider the distribution of hydrocarbons depicted in \ref{fig:excludedModel}, we recover unphysical (negative) values for certain loss tangents, which implies that the model is incorrect. This suggests that piranha cleaning is effective at removing fabrication related hydrocarbons from the sapphire surface.

\section{Oxide thickness after long BOE} \label{sec:longBOEThick}

In \cite{mclellan_chemical_2023}, we measure the oxide thickness of tantalum films under three surface conditions: native, BOE treated for 20 minutes, and BOE treated for 40 minutes. The technique used, variable energy XPS (VEXPS) requires a synchrotron light source, and so could not be replicated in our lab to measure the thickness of the 120 minute BOE treated surface. Instead, to estimate the total oxide thickness, we correlate XPS measurements done on our laboratory system to oxide thickness measurements from \cite{mclellan_chemical_2023}.

\begin{figure}[htbp]
    \centering
    \includegraphics[width=6.75in]{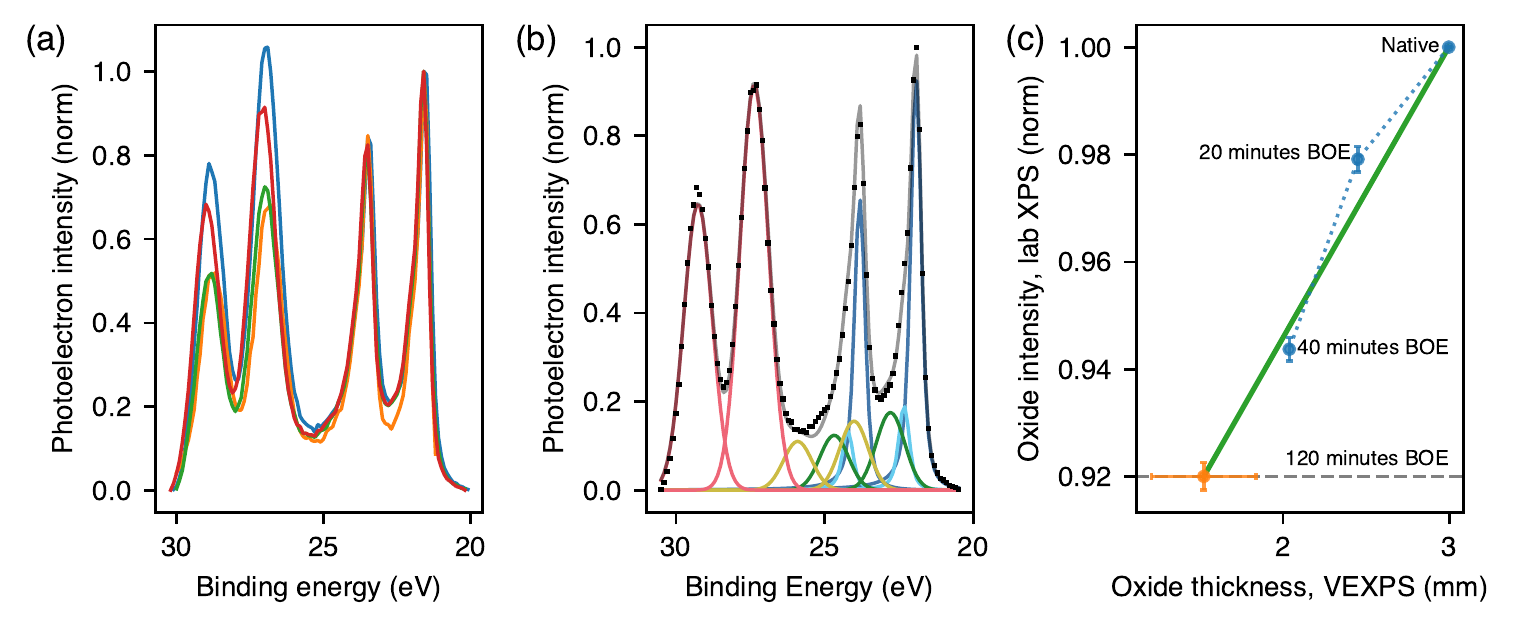}
    \caption{(a) Ta4f spectra for native (blue), 20 minute BOE treated (red), 40 minute BOE treated (green) and 120 minute BOE treated (orange) samples. Data are Shirley background corrected \cite{engelhard_introductory_2020} and normalized so the peak height of the Ta$_{7/2}$ metallic peak is unity. (b) Fit to the XPS peaks for the 20 minute BOE treated sample. The peaks used to fit the spectrum are doublets of Ta$^0$ (dark blue), Ta$^0_\textrm{int}$ (cyan), Ta$^{1+}$ (green), Ta$^{3+}$ (yellow), and Ta$^{5+}$ (pink). Ta$^0$ and Ta$^0_\textrm{int}$ peaks are fit with asymmetric Voigt profiles, others are fit with symmetric Gaussians. The lower binding energy peak in each doublet corresponds to the Ta$_{7/2}$ spin state and the higher to the Ta$_{7/2}$ spin state \cite{Moulder:1992}. (c) Correlation of oxide photoelectron intensity fit to our lab-based XPS data to the oxide thickness measured in VEXPS \cite{mclellan_chemical_2023}. Data (blue) is available from VEXPS for native, 20 minute BOE treated, and 40 minute BOE treated samples. Green is the best fit line to these data points, extrapolated to the oxide photoelectron intensity fraction of the 120 minute BOE treated sample (grey dashed line) to give an estimate of the oxide thickness after a 120 minute BOE treatment (orange). Photoelectron intensity fractions are normalized so that the native intensity fraction is unity.}
    \label{fig:sup_XPS120BOE}
\end{figure}

We measured the Ta4f spectrum for four samples with four surface treatments: native; and treated in BOE for 20, 40, and 120 minutes (Figure \ref{fig:sup_XPS120BOE}(a)). We subtracted a Shirley background from all spectra \cite{engelhard_introductory_2020} and normalized all data so that the metallic Ta$_{7/2}$ peak height is unity. Similar to our observations in Section \ref{sec:XPS}, we see a decrease in the photoelectron fraction from the Ta$^{5+}$ species with BOE treatment. We fit all Ta4f spectra with doublets associated with the Ta$^0$, Ta$^0_\textrm{int}$, Ta$^{1+}$, Ta$^{3+}$, and Ta$^{5+}$ states \cite{mclellan_chemical_2023, himpsel_core-level_1984}. The Ta$^0$ and Ta$^0_\textrm{int}$ peaks are all fit with asymmetric Voigt peaks, while the Ta$^{1+}$, Ta$^{3+}$, and Ta$^{5+}$ peaks are all fit with symmetric Gaussians.

We consider the total oxide thickness to be the sum of the Ta$^{1+}$, Ta$^{3+}$, and Ta$^{5+}$ species. We expect the fraction of the photoelectron intensity corresponding to these peaks to be proportional to their thickness with some unknown rate parameter. To find this rate parameter, we take the photoelectron intensity fraction of all oxide species for each sample and compare them to the oxide thicknesses measured in VEXPS (\ref{fig:sup_XPS120BOE}(c)). We find an approximate linear relationship between the photoelectron intensity fraction of the oxide and the measured oxide thickness in VEXPS. We extrapolate this line to the photoelectron intensity fraction of the 120 minute BOE treated device and find an estimated oxide thickness of 1.5~$\pm$~3~mm. The error in this estimate is dominated by the linear fit shown in \ref{fig:sup_XPS120BOE}(c).

\section{Correlations among other parameters}

\begin{figure}[htbp]
    \centering
    \includegraphics[width=6.75in]{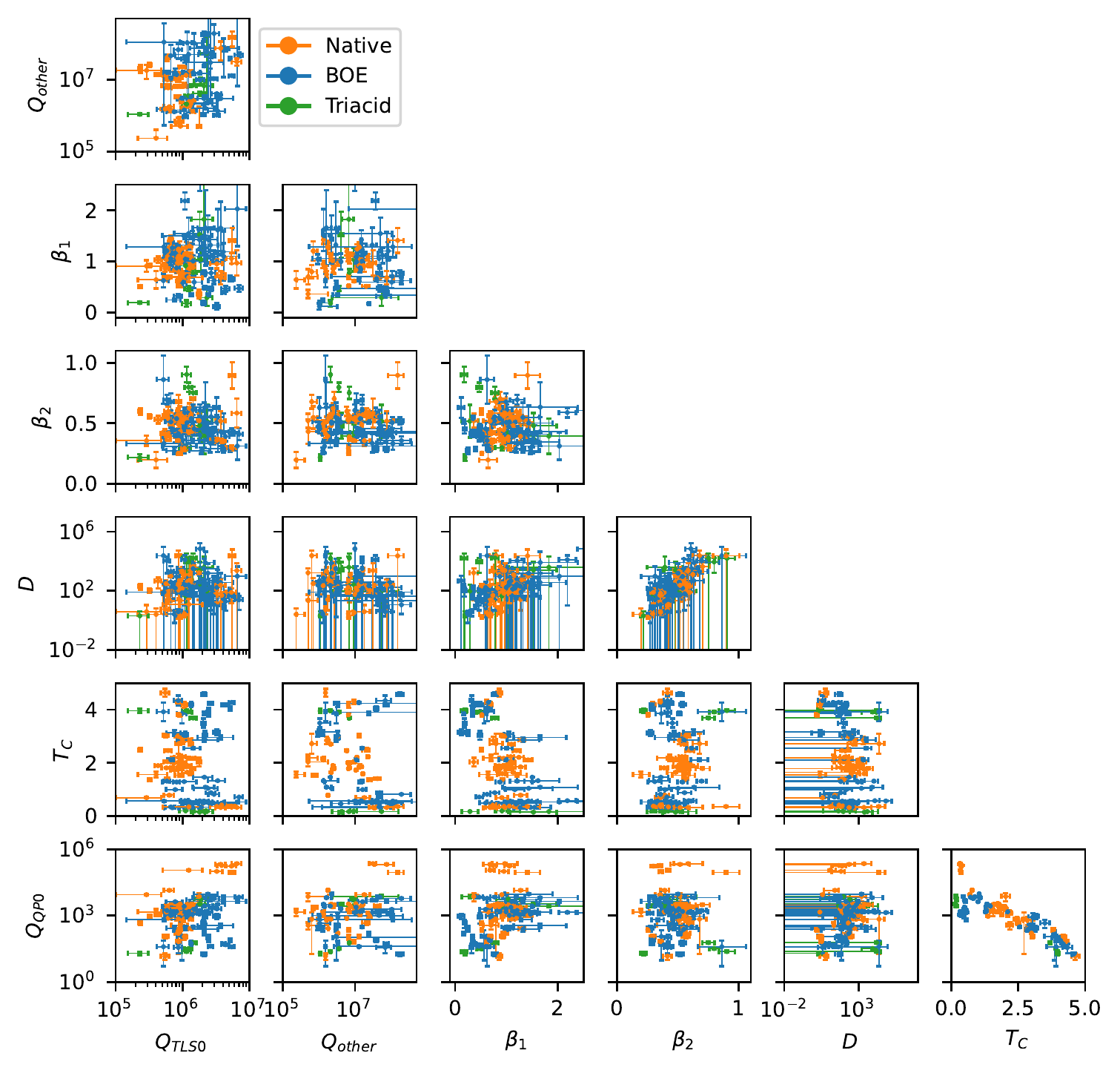}
    \caption{Correlations between all seven fitted parameters used in Equation 1, Equation 2, and Equation 3 in the main text. Lower uncertainty bounds in $Q_\textrm{other}$ are truncated to $Q_\textrm{TLS,0}$}
    \label{fig:sup_correlations}
\end{figure}

The model described in Equations 1-3 in the main text contains seven free fit parameters. To check that we can independently extract all seven parameters from our dataset, we plot each pair of fitted parameters in Figure \ref{fig:sup_correlations}. We see no correlations among parameters except between $T_c$ and $Q_{\textrm{QP,0}}$ and between $D$ and $\beta_2$.

The correlation between $T_c$ and $Q_{\textrm{QP,0}}$ is likely due to the limited amount of high temperature data that we recorded (Section \ref{sec:otherFits}). We conclude that, with our current dataset, we cannot quantitatively separate $T_c$ and $Q_{\textrm{QP,0}}$.

The correlation between $D$ and $\beta_2$ is an artifact of the parameterization of Equation 2. The correlation is a straight line on a log-linear plot, and therefore if we considered $D^{1/\beta2}$ as our fit parameter instead of $D$, we would see no correlation.

We note that no correlation is apparent between $Q_\textrm{TLS,0}$ and any of the other six parameters. Therefore we are able to meaningfully differentiate the linear absorption of our TLS bath from its saturation behavior, and from the effects of thermal quasiparticles and the constant loss parameterized by $Q_\textrm{other}$. In particular, we note that the fitted value of $T_c$ can range from 0.14~K to above 4~K, which can affect the available temperature range of the data (Section \ref{sec:otherFits}), but the fitted $Q_\textrm{TLS,0}$ value does not correlate with $T_c$.

\section{Additional variables}
\subsection{Effects of etch type on loss}
\begin{figure}[htbp]
    \centering
    \includegraphics[width=4in]{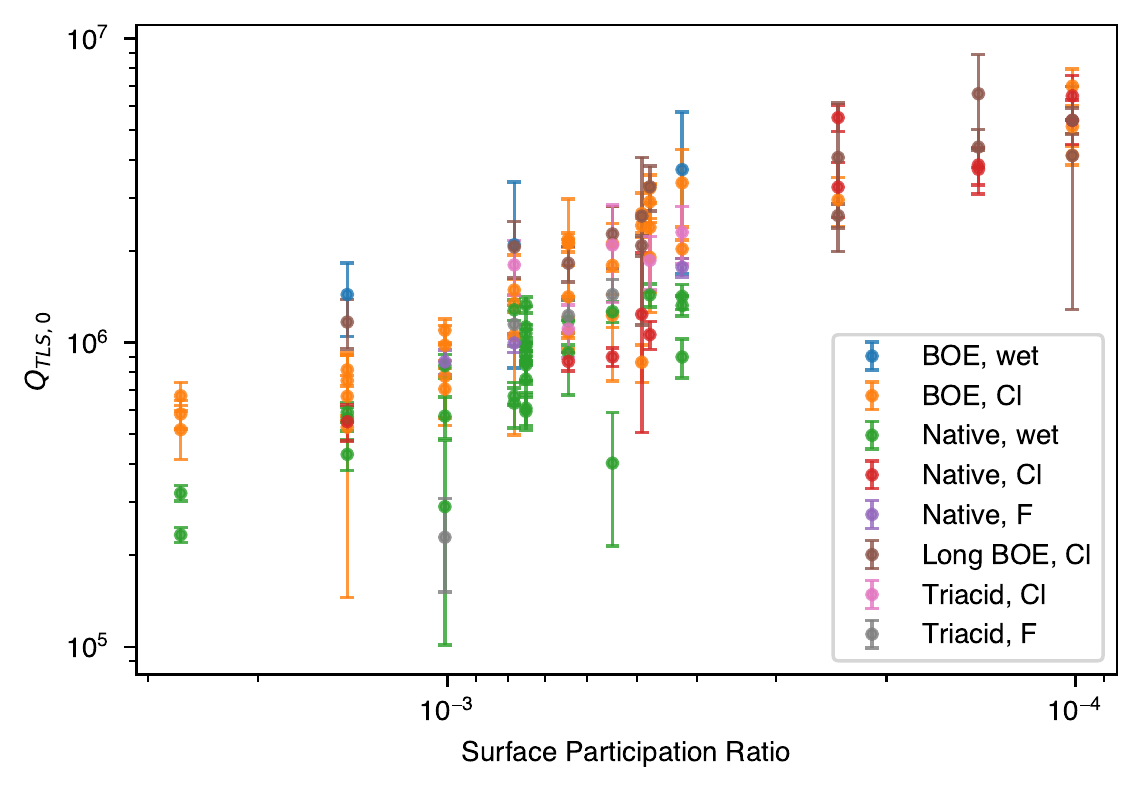}
    \caption{Dependence of $Q_{\textrm{TLS,0}}$ on SPR separated by etch type.}
    \label{fig:sup_etchComp}
\end{figure}

In addition to the surface treatments we discuss in the main text, we also vary the method of etching the tantalum, as described in Section \ref{sec:fab}. The plot of $Q_{\textrm{TLS,0}}$ vs $p_{\textrm{MS}}$ in Figure 3 of the main text is reproduced here, but with the data points further stratified by etch type (Figure \ref{fig:sup_etchComp}). It is possible that dry etching with either the Cl or F based recipe leads to additional surface damage and TLS loss or that different etch types create different edge qualities, however our data does not have enough statistical power to conclusively identify another source of loss arising from etch type.

\subsection{Annealing sapphire}

One possible source of TLS loss is the disordered sapphire surface. We explored sapphire annealing to interrogate the contribution of this surface. Prior to tantalum deposition, we processed some of our sapphire wafers to achieve a near atomically-flat surface with observable step edges to probe the impact on the metal-substrate and substrate-air losses. Atomic terraces have been previously observed on sapphire after high temperature annealing \cite{dwikusuma_study_2002}.

After cleaning the sapphire wafers in a 2:1 piranha bath and rinsing dry, we then treated the sapphire in 146\degree C sulphuric acid (SigmaAldrich catalogue number: 258105) for 20 minutes, then triply rinsed the wafer in de-ionized water, rinsed once in 2-propanol, and then blew it dry in N$_2$. Finally, we annealed the sapphire in an air furnace with a temperature ramp of 4.1\degree C/minute to 1100\degree C, and then held at 1100\degree C for one hour.

\begin{figure}[htbp]
    \centering
    \includegraphics[width=5in]{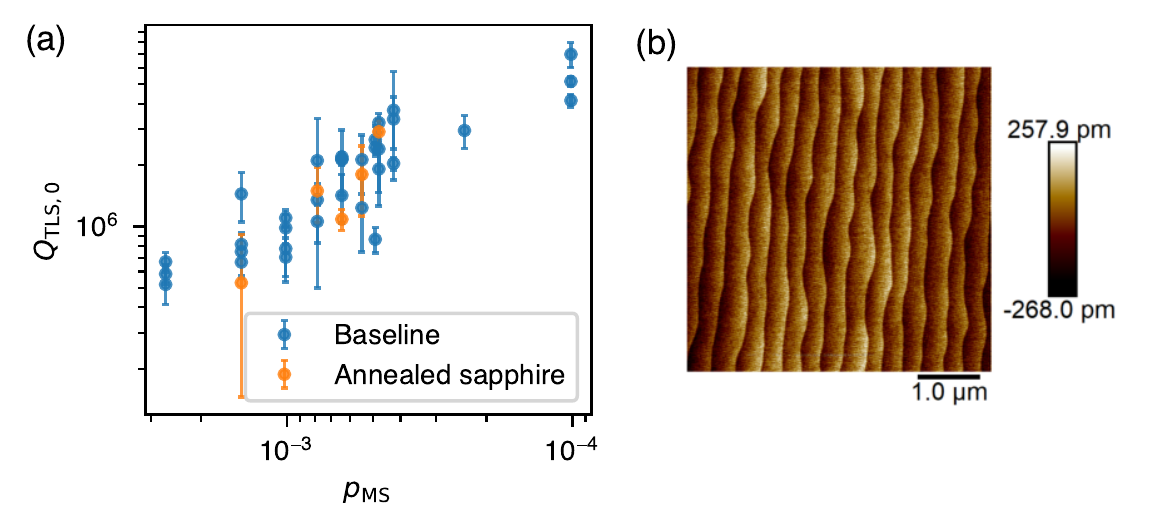}
    \caption{(a) Dependence of extracted $Q_\textrm{TLS,0}$ on SPR, separated into devices fabricated on annealed substrates and those fabricated on unannealed substrates. All devices were treated in BOE for 20 minutes. No significant difference in performance is seen. (b) AFM image of annealed sapphire surface. Scanned in 512 lines with a 1~Hz scan rate and a 7mm tip.}
    \label{fig:sup_anneal}
\end{figure}

We confirmed that we were able to achieve a flat surface by performing atomic force microscopy (AFM) on a sapphire sample prior to deposition (Figure \ref{fig:sup_anneal}(b)). The AFM used was a Bruker ICON3 with a 7~nm AFM tip. We observe discrete steps in height, each approximately 250~pm. This step size is on the order of a single lattice constant, and so we conclude we are observing atomic terraces.

We measured several resonators fabricated from films deposited on this annealed sapphire. All of these resonators were treated in BOE for 20 minutes. The dependence of $Q_\textrm{QTLS,0}$ on SPR for these resonators is not distinguishable from the dependence seen for other BOE treated resonators (Figure \ref{fig:sup_anneal}(a)). As there is no measurable effect, we conclude that the sapphire anneal does not affect TLSs that are limiting our devices. We note that all resonators that were measured on annealed sapphire had high SPRs, in the linear region in Figure \ref{fig:sup_anneal}(a), and thus are not sensitive to changes in bulk loss. An interesting avenue for future exploration would be to see if high temperature annealing can change the bulk loss in sapphire.

\subsection{Packaging}

We used three types of package in our experiments. The first was a copper ``puck and penny" assembly, the second was the commercially available QCage.24 from QDevil, and the third was a modified version of the QCage.24 with an aluminum flashed coating on the surfaces of the package which face the device. 

\begin{figure}[htbp]
    \centering
    \includegraphics[width=4.5in]{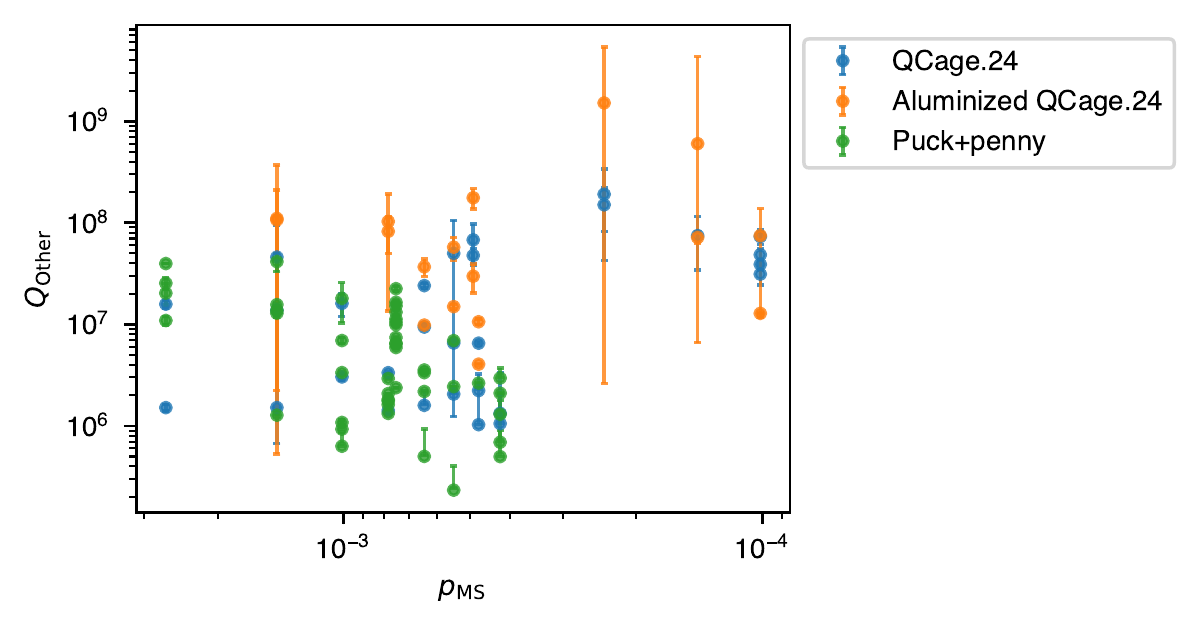}
    \caption{Dependence of extracted $Q_\textrm{other}$ on SPR, separated into devices packaged into the puck and penny assembly, the QCage.24 with bare copper surfaces, and the QCage.24 with aluminum flashed surfaces. Lower error bars are truncated to the value of $Q_\textrm{TLS,0}$.}
    \label{fig:sup_package}
\end{figure}

We compare the measured values of $Q_\textrm{other}$ achieved for devices in each of the three packages in Figure \ref{fig:sup_package}. We find that higher values of $Q_\textrm{other}$ can be achieved for the QCage.24 package, with the highest values achieved with the aluminum flashing. Nine total devices packaged in the QCage.24 and aluminum flashed QCage.24 were excluded from Figure \ref{fig:sup_package}, as $Q_\textrm{other}$ was too large relative to the measured values of $Q_\textrm{int}$ to be able to be fit confidently.

We conclude that, in some cases, $Q_\textrm{other}$ is limited by packaging loss which are present on the puck and penny assembly, but not present on the QCage.24. Based on the fact that the highest values of $Q_\textrm{other}$ are found with aluminum flashing on the inside of the QCage.24, we further conclude that in the QCage.24, the electric field of the modes of our devices have a non-negligible overlap with the packaging material. The improvement in $Q_\textrm{other}$ is achieved by having the nearest surface of the package be a superconducting metal.

\subsection{Surface morphology}

\begin{figure}[htbp]
    \centering
    \includegraphics[width=6.75in]{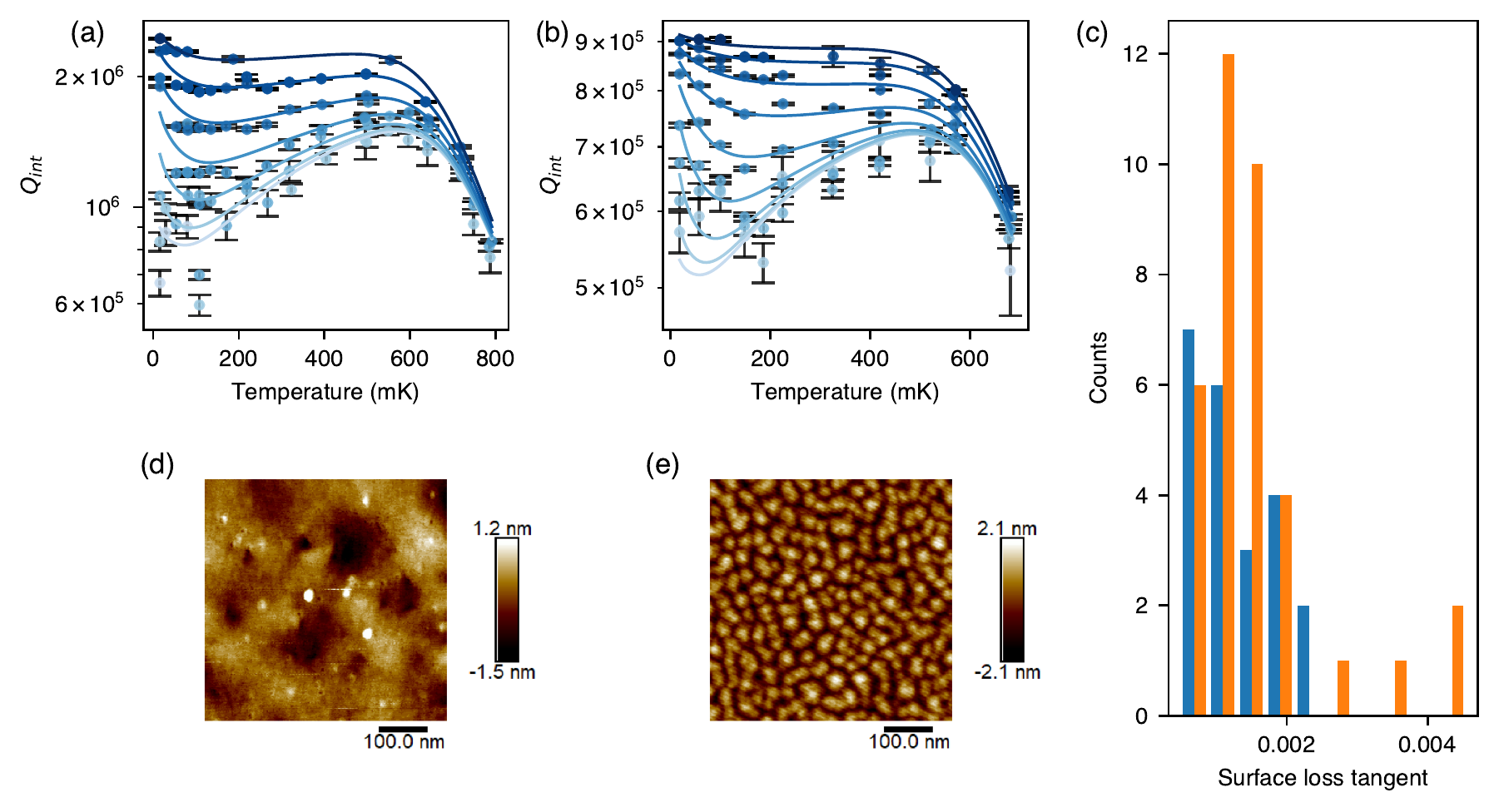}
    \caption{Effect of surface morphology on device performance. Tantalum films used in our experiment were deposited by our group or by Star Cryoelectronics, and films from the two different sources show qualitatively different surface morphologies. (a-b) Example temperature sweeps from devices fabricated tantalum deposited by our group (a) or Star Cryoelectronics (b). Both devices were BOE treated and have surface participation ratios of approximately 10$^{-3}$. The color represents input power, with the darkest shade being the highest power. The spacing between powers is 10~dB. (c) Histogram of surface loss tangents from devices fabricated on films deposited by our group (blue) and Star Cryoelectronics (orange). Only devices with a BOE treatment are included. (d-e) Example atomic force microscopy images showing surface morphology on a film deposited by our group (d) and by Star Cryoelectronics (e). The color scale represents depth.}
    \label{fig:sup_morphology}
\end{figure}

Our tantalum films were deposited both by our group and by Star Cryoelectronics. Both sources showed a body-centered cubic $\alpha$-Ta phase with majority $\langle 111\rangle$ orientation when measured with an X-ray diffractometer, however, the surface morphology as measured with atomic force microscopy (AFM) is qualitatively different. Figure \ref{fig:sup_morphology}(d) and Figure \ref{fig:sup_morphology}(e) show AFM images (Bruker ICON3) taken on the tantalum surface of films deposited by our group and Star Cryoelectonics, respectively. 

We see no qualitative difference in the temperature sweep data between the two types of films (Figure \ref{fig:sup_morphology}(a-b)). We compare the fitted values of the surface loss tangent from devices with the same surface treatment fabricated on films from the two sources, and see no significant difference (Figure \ref{fig:sup_morphology}). We conclude that any losses associated with this variation in observed surface morphology difference do not limit device performance.

\subsection{Rapid thermal annealing}

With XPS, we can observe a shoulder peak at approximately 0.4~eV higher binding energy than the metallic tantalum peaks. Peaks in this location have been observed in \cite{himpsel_core-level_1984}, in which they are attributed to the closest layer of tantalum metal atoms to the oxide and have a differing coordination number to those in the bulk. A plausible hypothesis for a location of TLSs is in this interfacial tantalum layer.

Rapid thermal annealing (RTA) is used in semiconductor processing to increase the orderng of interfacial layers in thin films \cite{singh_rapid_1988}, and has been shown to have an effect on tantalum oxide thin films \cite{ezhilvalavan_short-duration_1999}. We used (RTA) to change the metal-oxide interface. Our process consisted of a ramp to 800\degree C in 30 seconds and holding at 800\degree C for a further 30 seconds. The process was completed in a Thermo Scientific Lindberg/Blue M furnace (PN: STF55433C-1).

\begin{figure}[htbp]
    \centering
    \includegraphics[width=6.75in]{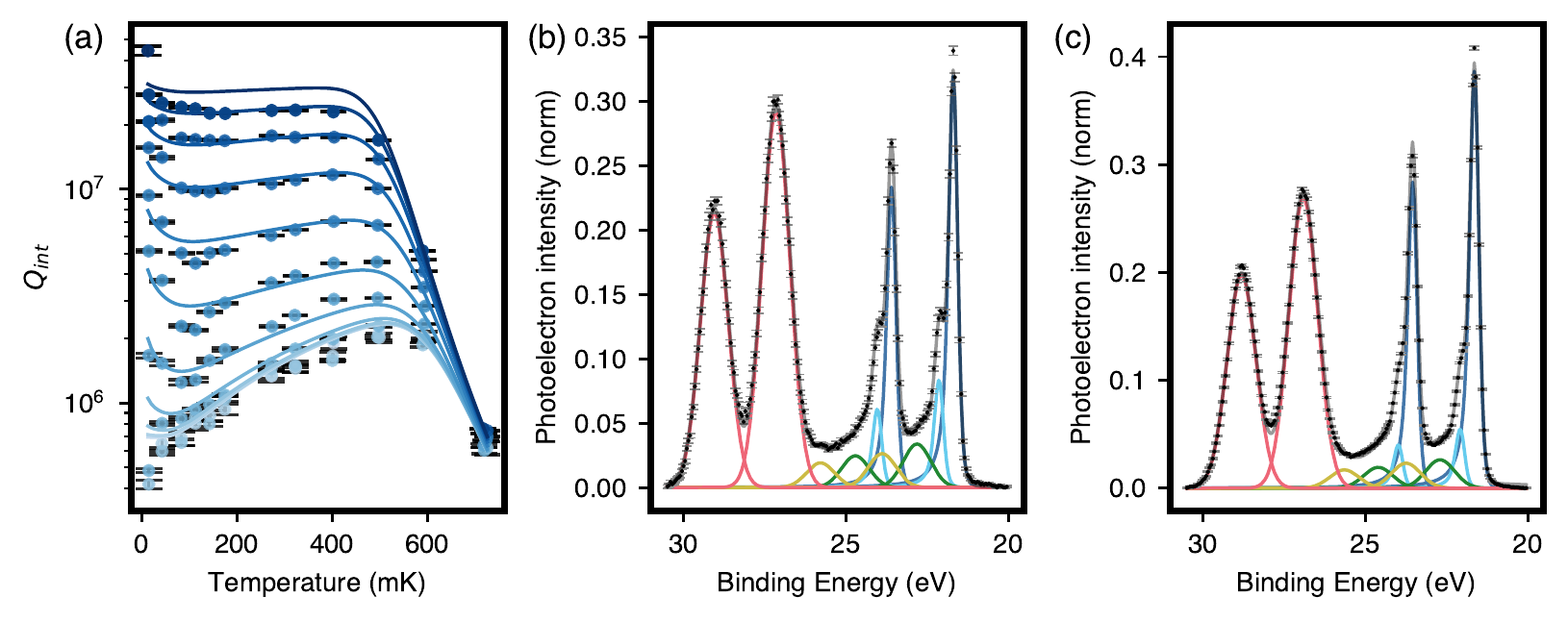}
    \caption{(a) Results from temperature sweep fitted to a device treated with a rapid thermal anneal followed by a BOE treatment. The calculated surface participation ratio is approximately 2.6$\times 10^{-3}$. The fitted value of $Q_\textrm{TLS,0}$ is (6.97 $\pm$ 0.36)$\times 10^5$. (b-c) XPS data and fits for the Ta4f peaks performed on native films without (b) and with (c) the RTA process. All data are Shirley background corrected \cite{engelhard_introductory_2020} and normalized so the total intensity for the spectrum is unity. The peaks used to fit the spectrum are doublets of Ta$^0$ (dark blue), Ta$^0_\textrm{int}$ (cyan), Ta$^{1+}$ (green), Ta$^{3+}$ (yellow), and Ta$^{5+}$ (pink). Ta$^0$ and Ta$^0_\textrm{int}$ peaks are fit with asymmetric Voigt profiles, others are fit with symmetric Gaussians. The lower binding energy peak in each doublet corresponds to the Ta$_{7/2}$ spin state and the higher to the Ta$_{7/2}$ spin state \cite{Moulder:1992}. There is an approximately 20\% decrease in the fitted intensity of the Ta$^0_\textrm{int}$ peaks. Data taken at the Spectroscopy Soft and Tender 2 (SST-2) endstation at the National Synchrotron Light Source II at Brookhaven National Lab with X-ray energy 2000~eV. Data were collected with the same methodology described in \cite{mclellan_chemical_2023}}
    \label{fig:sup_RTA}
\end{figure}

We performed XPS on native samples with and without the RTA process, observed a decrease of approximately 20\% in the fitted intensity of the interfacial tantalum shoulder peak (Figure \ref{fig:sup_RTA}(b-c)). We performed this RTA process on a resonator chip and measured a temperature sweep (Figure \ref{fig:sup_RTA}(a)). We observe no qualitative difference between the temperature sweep data on this device and data from temperature sweeps on devices without the RTA process. For the device fabricated on the film with the RTA process, the fitted $Q_\textrm{TLS,0}$ is (6.97 $\pm$ 0.36)$\times 10^5$ with a SPR of  2.6$\times 10^{-3}$. 

We measured three other devices with an SPR of of 2.6$\times 10^{-3}$ that were BOE treated, and we extracted a mean Q$_\textrm{TLS,0}$ of (6.0 $\pm$ 0.4)$\times 10^5$. The device which underwent RTA has a $Q_\textrm{TLS,0}$ over 2$\sigma$ higher than the mean $Q_\textrm{TLS,0}$ for the control devices, we conclude that we may have seen a significant performance difference due to RTA. However, given that we only measured one RTA device, we cannot rule out the possibility that there would be a change in the extracted loss tangent of a family of RTA processed devices. 

\section{Examples of temperature fits} \label{sec:otherFits}

\begin{figure}[htbp]
    \centering
    \includegraphics[width=6in]{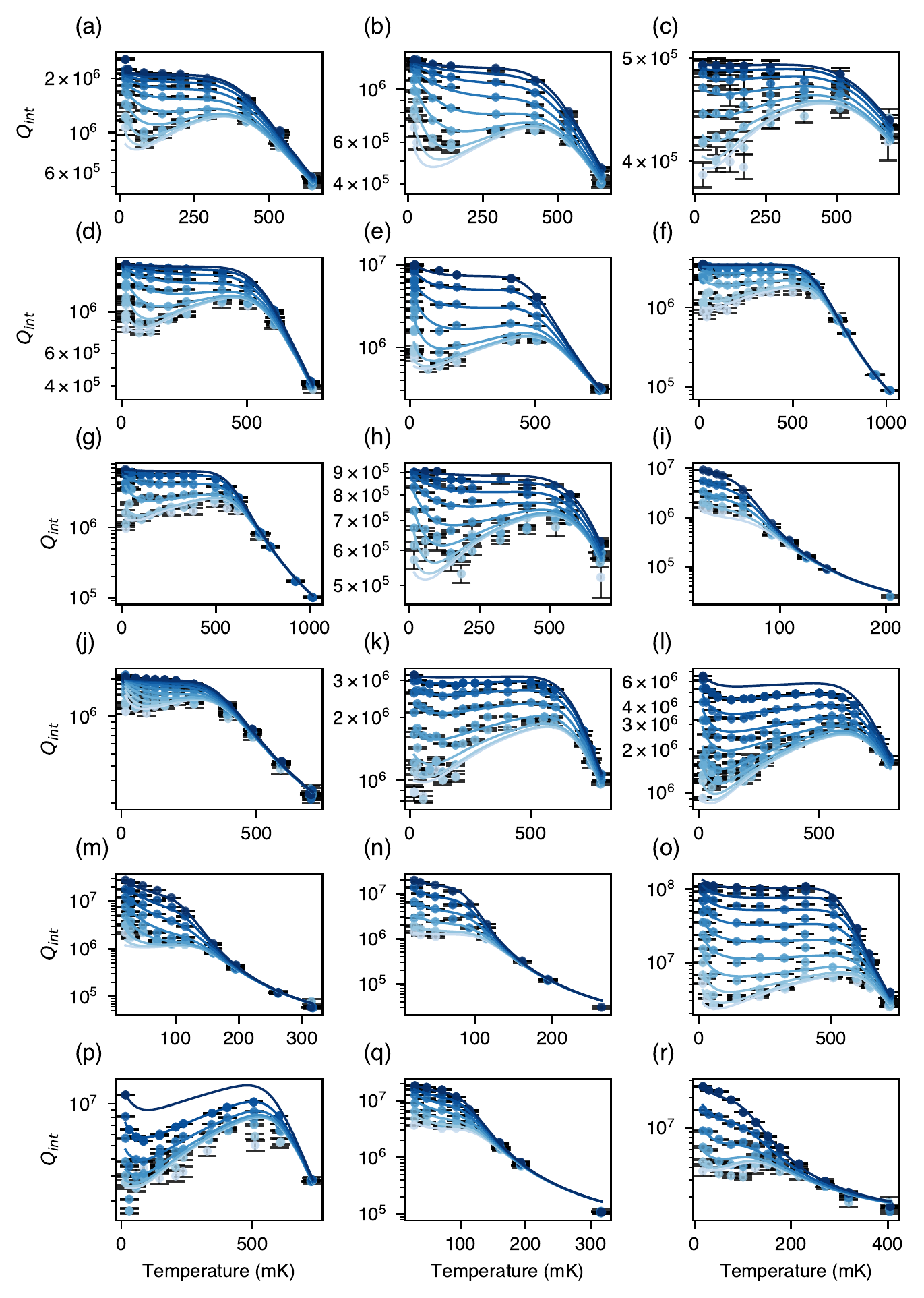}
    \caption{Examples of representative fits to temperature sweep data. Data are taken from a variety of LE and CPW devices, as well as from native, BOE treated, long BOE treated, and triacid treated surfaces. Colors indicate circulating power in the feedline, with the darkest shade representing the highest power the lightest shade the lowest power. All traces spaced 10~dB apart.}
    \label{fig:sup_otherPlots}
\end{figure}

Additional examples of fits to temperature sweep data are shown in Figure \ref{fig:sup_otherPlots}. The SPR, surface treatment, resonator type, and packaging corresponding to each resonator is given in Table \ref{table:sup_otherPlotsTable}.

We note that some devices (Figure \ref{fig:sup_otherPlots}(g,i,j,m,n,q,r)) begin to be dominated by equilibrium quasiparticles at a lower temperature, which we attribute to a minority phase of $\beta$-Ta that was below the detectable limit for our X-ray diffractometer setup. We also note that some resonators show a small range of $Q_\textrm{int}$ (Figure \ref{fig:sup_otherPlots}(f,g,j,q)), indicating that the TLS loss and non-saturable loss mechanisms, parameterized by $Q_\textrm{other}$, are becoming comparable. Lastly, we note that some resonators (Figure \ref{fig:sup_otherPlots}(p,r)) do not show evidence of a power- and temperature-independent loss mechanism in the ranges of $Q_\textrm{int}$ shown. We were unable to fit $Q_\textrm{other}$ to these devices.

\begin{table}[htbp]
\caption{Device and measurement parameters corresponding to data shown in Figure \ref{fig:sup_otherPlots}. Surfaces are native (N), triacid (T), BOE (BOE) or long BOE (LB). Packaging is either puck and penny (P), QCage.24 (Q), or QCage.24 with aluminum flashing (QAl). Etch type is wet (W), dry chlorine based (Cl), or dry fluorine based (F).}
\centering
\setlength{\tabcolsep}{2pt}
\begin{tabular}{|l l l l l l l l l l l l l l l l l l l|}
    \hline
    Subplot & (a) & (b) & (c) & (d) & (e) & (f) & (g) & (h) & (i) & (j) & (k) & (l) & (m) & (n) & (o) & (p)& (q) & (r) \\
    \hline
    SPR ($\times 10^{-4}$) & 6.4 & 4.2 & 4.2 & 7.8 & 14.4 & 6.4 & 5.4 & 10.1& 10.1 & 5.5 & 7.8& 6.4& 6.4&13.3& 4.9&4.9&4.9&1.4 \\
    Surface     & N    & N   & N & N & N & T & T & BOE & BOE &BOE & BOE & N & BOE&LB &LB&SB&LB&N \\
    Device type            & CPW  & CPW & CPW & CPW & CPW & CPW & CPW & CPW & CPW & CPW & CPW & CPW& CPW&CPW&LE&LE&LE&LE \\
    Packaging              & P    & P & P & P & P & P & P & P & Q & Q & Q&Q&QAl&QAl&QAl&Q&QAl&Q \\
    Etch & W &W & F & W & W & F & Cl & Cl & Cl & Cl & Cl & Cl & Cl & Cl & Cl & Cl & Cl & Cl \\
    \hline
\end{tabular}
\label{table:sup_otherPlotsTable}
\end{table}

\clearpage
\bibliography{resonatorPaperBibliography} 